\documentclass[iop]{emulateapj}
\usepackage{color}
\usepackage{psfig}

\slugcomment{(Draft \today)}

\gdef\imacs{{\it IMACS}}

\def\etal{\hbox{et al.}}

\gdef\ltsima{$\scriptscriptstyle \; \buildrel < \over \sim \;$}
\gdef\simlt{\lower.3ex\hbox{\ltsima}}

\gdef\gtsima{$\scriptscriptstyle \; \buildrel > \over \sim \;$}
\gdef\simgt{\lower.3ex\hbox{\gtsima}}

\gdef\about{\raise.3ex\hbox{$\scriptscriptstyle \sim $}}

\def\gs{\mathrel{\raise0.35ex\hbox{$\scriptstyle >$}\kern-0.6em 
\lower0.40ex\hbox{{$\scriptstyle \sim$}}}}
\def\ls{\mathrel{\raise0.35ex\hbox{$\scriptstyle <$}\kern-0.6em 
\lower0.40ex\hbox{{$\scriptstyle \sim$}}}}

\def\etal{\hbox{et al.}}

\def\Msun{\rm{\hbox{$M_{\odot}$}}}				
\def\MsunYr{\rm{\hbox{M$_{\odot}$} yr$^{-1}$}}           
\def\ang{\hbox{$\,$\AA}}

\def\24m{\hbox{24\,$\micron$}$\,$}
\def\10-18{\hbox{$\times~10^{-18}$}}

\def\Tobs{\hbox{\emph{T$_{obs}$}}}
\def\Tmax{\hbox{\emph{Tmax}}}
\def\T0{{$t_0$}}

\def\R200{\hbox{$R_{200}$}}

\slugcomment{Submitted to the Astrophysical Journal: 2016 July 5; accepted: 2016 October 21}

\shortauthors{Dressler \etal\ }
\shorttitle{Demonstrating Diversity in Star-Formation Histories with the CSI Survey}

\begin{document}

\title{Demonstrating Diversity in Star-Formation Histories with the CSI Survey
\footnote{T\lowercase{his paper includes data gathered with the 6.5 meter \uppercase{M}agellan \uppercase{T}elescopes 
located at \uppercase{L}as \uppercase{C}ampanas \uppercase{O}bservatory, \uppercase{C}hile.}}}

\author{
Alan Dressler\altaffilmark{1},
Daniel D. Kelson\altaffilmark{1},
Louis E. Abramson\altaffilmark{2},
Michael D.~Gladders\altaffilmark{3,4},
Augustus Oemler, Jr.\altaffilmark{1},
Bianca M.~Poggianti\altaffilmark{5},
John S. Mulchaey\altaffilmark{1},
Benedetta Vulcani\altaffilmark{6},
Stephen A. Shectman\altaffilmark{1},
Rik J. Williams\altaffilmark{7},
Patrick J. McCarthy\altaffilmark{1}
}

\altaffiltext{1}{The Observatories of the Carnegie Institution for Science, 813 Santa Barbara St., Pasadena, CA 91101, USA}
\altaffiltext{2}{Department of Physics and Astronomy, University of California, Los Angeles, 475 Portola Place, Los Angeles, CA 90095-1547, USA }
\altaffiltext{3}{Department of Astronomy and Astrophysics, University of Chicago, 5640 S. Ellis Ave., Chicago, IL 60637, USA}
\altaffiltext{4}{Kavli Institute for Cosmological Physics, University of Chicago, 5640 S. Ellis Ave., Chicago, IL 60637, USA}
 \altaffiltext{5}{INAF-Astronomical Observatory of Padova, Italy}
\altaffiltext{6}{School of Physics, The University of Melbourne, VIC 3010, Australia} 
\altaffiltext{7}{United States Agency for International Development, 1300 Pennsylvania Ave., NW, Washington DC 20004-3002, USA}

\begin{abstract}

We present coarse but robust star-formation histories (SFHs) derived from spectrophotometric data of the Carnegie-Spitzer-IMACS Survey, 
for 22,494 galaxies at $0.3<z<0.9$ with stellar masses of 10$^9$\,\Msun\ to 10$^{12}$\,\Msun.  Our study moves beyond ``average" SFHs 
and distribution functions of specific star-formation rates (sSFRs) to individually measured SFHs for tens of thousands of galaxies.  By 
comparing star-formation rates (SFRs) with timescales of $10^{10}, 10^9$, and $10^8$ years, we find a wide \emph{diversity} of SFHs: 
``old galaxies" that formed most or all of their stars early; galaxies that formed stars with declining or constant  SFRs over a Hubble time, 
and genuinely ``young galaxies" that formed most of their stars since $z=1$.  This sequence is one of decreasing stellar mass, but, remarkably, 
each type is found over a mass range of a factor of 10.  Conversely, galaxies at any given mass follow a wide range of SFHs, leading us to 
conclude that (1) halo mass does not uniquely determine SFHs, (2) there is no ``typical'' evolutionary track, and (3) ``abundance matching" has 
limitations as a tool for inferring physics.  Our observations imply that SFHs are set at an early epoch, and that---for most galaxies---the decline 
and cessation of star formation occurs over a Hubble time, without distinct ``quenching" events.  SFH diversity is inconsistent with models 
where galaxy mass, at any given epoch, grows simply along relations between SFR and stellar mass, but is consistent with a two-parameter 
lognormal form, lending credence to this model from a new and independent perspective.

\end{abstract}


\section{Introduction: Star-Formation Histories---Conformal, or Diverse? }

Large data samples of galaxy photometry are now available from the present epoch back to $\sim$1 Gyr after the Big Bang.
Hundreds of studies have described and analyzed these data in terms of luminosity, mass, and structural evolution, relying
on trends between such quantities that assume a considerable uniformity of the growth of stellar populations and, by
implication, of dark-matter halos.  

Several considerations, including \emph{N}-body simulations, the well-populated trend of cosmic SFR density as a function of redshift 
(Lanzetta, Wolfe, \& Turnshek 1995, Lilly \etal\ 1996,  Pei \& Fall 1995, Madau \& Dickenson 2014), and the characterization 
of the controversially-named ``star-formation main sequence" (SFMS: Noeske \etal\ 2007; Whitaker \etal\ 2012)---showing a 
correlation between stellar mass and SFR at every epoch---have guided many studies in the crafting of mean evolutionary tracks, 
whose nature might lend insight into the phenomena driving the evolution of  individual galaxies (e.g., Whitaker \etal\ 2014).

However, it has been difficult to use such data to go beyond average properties and average evolution, to measuring the star 
formation histories (SFHs) of individual or classes of galaxies.  Such data would inform to what degree galaxies follow similar 
growth histories, offset or scaled in cosmic time, or whether there is a genuine diversity in SFHs that is non-conformal.  That is, 
they could demonstrate whether \emph{measured} SFHs fail to conform to scenarios wherein the evolution of scaling
 relations {\it controls} (as opposed to \emph{reflects}) galaxy growth (Peng \etal\ 2010; Leitner \etal\ 2012; Behroozi  
\etal\ 2013), by crossing or not appearing as offset/scaled versions of each other in mass or time. Such data could also determine 
if the significant scatter in the SFMS represents fundamental, long-term diversity in SFHs---generating an illusion of uniform 
growth patterns---or only a distracting perturbation to physically informative ``average tracks.''

The fundamental problem is that the available data, including integrated mass functions over most of cosmic time, are unable 
to uniquely ``connect the dots" between one epoch and another: the galaxies at later epochs are not necessarily the decendants 
of earlier galaxies observed to follow a similar trend.  Abramson \etal\ (2016) have in particular emphasized the ambiguity of the 
presently available diagnostics by showing that models in which galaxy growth is conformal over mass and those that show 
great diversity of SFHs are both able to pass the observational tests that the present data provide.  The promise of ``average SFHs" to 
elucidate important physical processes in galaxy evolution has arguably blinded us to the possibility that more physics will be learned 
from their diversity than from their sameness.

This ambiguity can be broken by measuring the SFHs of individual galaxies, but this has proven very difficult to do, particularly 
because---from our vantage point in the local universe---stellar populations more than 2 Gyr old are essentially indistinguishable 
from each other.  For this reason, ``population-synthesis models'' have been ineffective in describing the build-up of stellar mass even 
in relatively nearby, well-observed galaxies outside the Local Group, although considerable progress is coming from \emph{Hubble Space 
Telescope}-WFC3 observations, for example, the study of resolved  stellar populations in M31 (Dalcanton \etal\ 2012). Applying the 
population-synthesis technique at higher redshift would allow a better resolution of the ages of older populations, but spectral 
observations with sufficient resolution and depth are costly.

One might argue that we already know that SFHs are diverse. The iconic elliptical galaxy NGC3379 in the Leo Group 
likely formed most of its stars before $z=2$, while common galaxies of the same mass, for example, the Milky Way and 
Andromeda, have been puttering along for the full age of the universe.  However, the proponents of the ``conformal" model 
suggest that both NGC3379 and these starforming spirals are following the same growth history, but that star formation was 
\emph{quenched} in the former by some mechanism that rendered inaccessible the considerable amount of gas still 
in its vicitiny.  Some method of forcing long cooling times by heating the gas is the likely process.  Suggested mechanisms include  
a transition from cold accretion to a hot halo driven by dark-matter halo growth (Kere{\v s} \etal\ 2005, Dekel and Birnboim 2006), 
heating from an active galactic nucleus (AGN) or supernovae feedback  (Voit \etal\ 2015b), or the transition from low-entropy to high-entropy gas that is unable 
to cool onto the host galaxy (Voit \etal\ 2015a; Voit \etal\ 2015c).  

For galaxies like NGC3379 ``quenching'' could have been rapid---likely a gigayear or less.  But what, then, of galaxies like the 
Milky Way and Andromeda?  Both could have been forming stars consistently at a few \MsunYr\ for most of a Hubble time, or 
they could have had SFRs of $\sim$10 \MsunYr\ in the first few billion years of their history, falling steadily since $z\sim1$ to 
reach their present rates of $\sim$1\,\MsunYr.\footnote{Basic observational data for our Galaxy cannot  distinguish between 
the two, if one is willing to accept that the present trickle of star formation is a temporary lull.}   It is clear from recent studies 
measuring ultraviolet flux and Balmer absorption lines (Schawinski \etal\ 2014; Dressler \& Abramson 2015, p140; Vulcani \etal\ 2015a) 
that galaxies traversing the color space between the ``blue cloud'' and the ``red sequence'' are mostly doing so over billions of years, not 
in a $\ls$1\,Gyr timescale following an abrupt termination of star formation.   Is such a slow decline a ``quenching?''  Has it been 
triggered by an event, such as an AGN or intense feedback from star formation, or is it instead nothing more than a slow exhaustion 
of gas suitable for star formation?  Rather than ``slow quenching," as it is now being called (e.g., Barro \etal\ 2016), perhaps this
is just the normal course of galaxy evolution.  Both pictures have been valid, given the tools we have used to describe galaxy 
evolution.

Gladders \etal\ (2013, hereinafter G13) developed the fast-clock/slow-clock model by assuming that galaxy SFHs are lognormal 
in cosmic time.  This work was motivated by two observations: 

\begin{itemize}
\item{Oemler \etal\ (2013b, hereinafter O13) studied distribution functions of specific star formation rates (sSFRs) for a 
sample of galaxies with redshifts $0.0 < z < 0.8$ and found that an increasing fraction had rising SFRs earlier in this epoch.  
The presence of such galaxies, whose abundance increasse steadily from essentially zero today to $\sim$20\% at $z\sim1$, 
obviates the ``$\tau$-model," in which the most aggressive SFH is  constant in time (Tinsley 1972)}

\item{G13, noted that the evolution of the cosmic star formation rate density (SFRD, the `Madau-Lilly' Diagram) from the present 
day back to $z\sim6$  is very well fit by a lognormal in time with two parameters, \T0---the half-mass time in the production of 
stellar mass, and $\tau$---the width of the lognormal.  

\begin{equation}
\color{black}
{\rm SFR} \propto \frac{1}{\sqrt{2\pi\tau^2}} \frac{\exp\left[-\frac{(\ln t - t_0)^2}{2\tau^2}\right]}{t}
\color{black}
\end{equation}

G13 adopted this as a parameterization of SFHs of \emph{individual} galaxies, and showed an existence proof that the distribution 
of sSFRs for 2094 present-epoch galaxies, and the SFRD could be simulataneously fit by the 
sum of 2094 lognormal SFHs.  Moreover, this model described the sSFR distributions at $z=0.2, 0.4$ from the \emph{ICBS} 
survey (Oemler \etal\ 2013a) and sSFR distributions for samples at $z=0.6, 0.8$ from data of the \emph{AEGIS} 
survey (Noeske \etal\ 2007).  This is the model that Abramson (2015) and Abramson \etal\ (2015, 2016) have found to be 
successful in fitting a variety of other data, including mass functions and the SFMS (zero-point, slope, and scatter) back to 
$z\sim2$, and the zero-point and slope back to $z\sim8$.}

\end{itemize}

Encouraging as these and other results may be, these tests are incapable of distinguishing between conformity and diversity 
in SFHs.\footnote{Another non-conformal approach, the stochastic SFH model explored by Kelson (2014b), is also able to 
pass a wide range of observational tests.}  Specifically, the work on lognormal SFHs based on the model in G13 demonstrates 
only that a large set of lognormal SFHs can be constructed that reproduces the available data well.  But, the individual SFHs in 
this model cannot be tagged to specific galaxies: it is only the distribution function of the \T0\ and $\tau$ parameters, not the 
assignment to individual galaxies, that is robust in this approach.


\begin{figure*}[t]

\centerline{
\includegraphics[width=7.0in, angle=0]{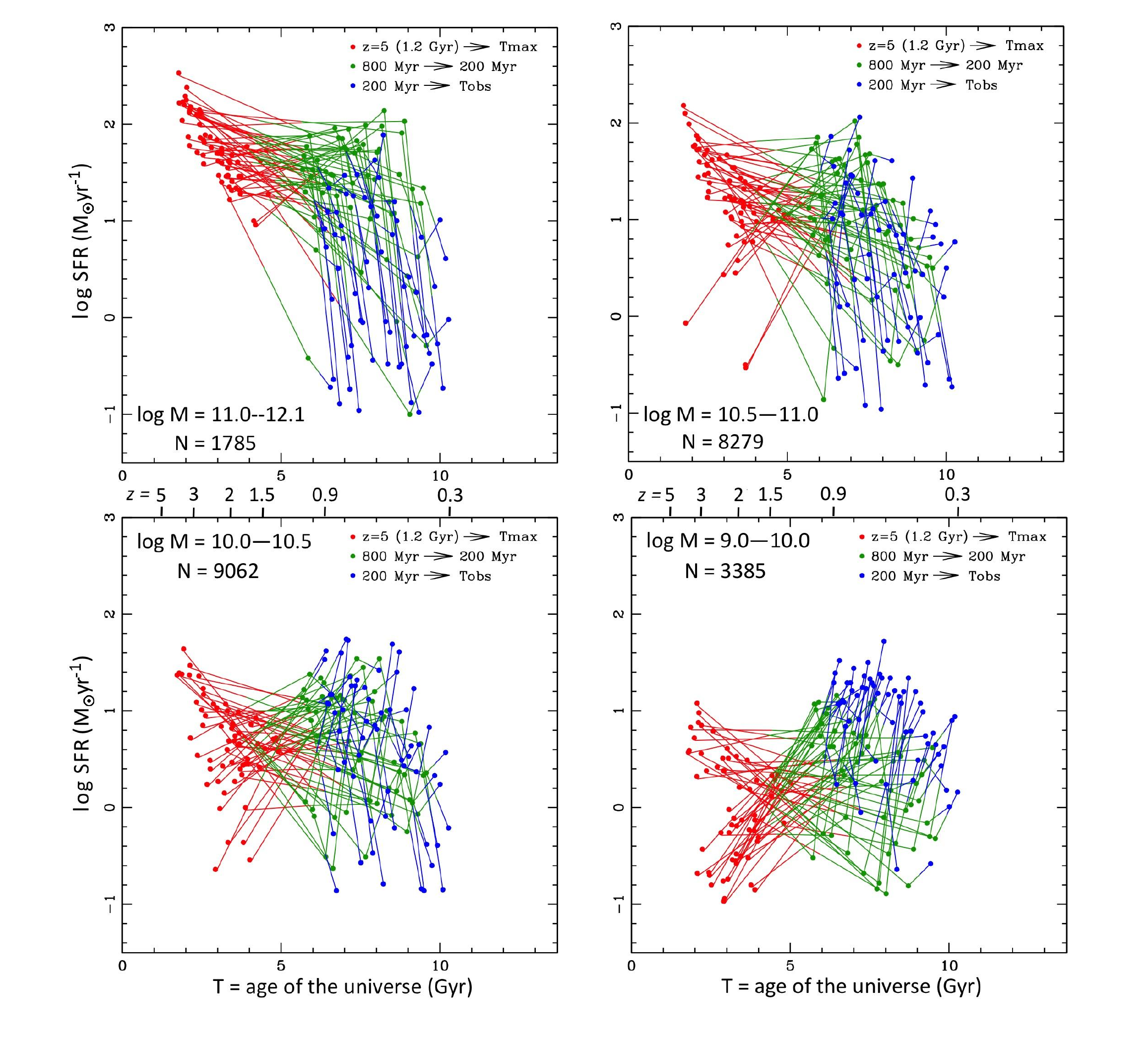}
}

\caption{Star-formation histories from the CSI Survey, plotted as log SFR in \MsunYr\ against cosmic time \emph{T}, the age of the 
universe.  The four panels are for different ranges in total stellar mass: (upper left) log M$_*$ = 11.0--12.1; (upper right) 
log M$_*$ = 10.5--11.0; (lower left) log M$_*$ = 10.0--10.5; (lower right) log M$_*$ = 9.0--10.0.  Each panel shows SFHs 
characterized by log SFRs at three epochs, for 60 randomly selected galaxies (in steps of 0.01 in redshift), out of the thousands that 
span the sample's redshift range,  $0.3<z<0.9$.  Each SFH has three measurements of SFR: (SFR1, red dots) $z=5$ (T=1.2 Gyr) to 1 Gyr 
before \Tobs\ (star formation not normally filling the full time); (SFR2, green dots) 1 Gyr to 200 Myr before \Tobs; and (SFR3, blue dots) 
200 Myr to \Tobs.  The dots are placed at the appropriate age of the universe for the mean of each star-formation epoch; for 
example, the green  dot of SFR2 is placed at T = \Tobs - 600 Myr    Typical 1$\sigma$ errors in log SFR are a 0.17, 0.27, and 
0.22 for SFR1, SFR2, SFR3, respectively.  As described in the text, the progression from the highest-mass (upper left) to the lowest-mass 
(lower right) galaxies shows a clear trend of SFHs that start early, with high SFRs that soon decline, toward an increasing fraction of 
galaxies with little or no early star formation that we refer to (following O13) as ``young galaxies" (not just the frosting, but the cake!).  
The latter dominate the SFHs in the lower right panel of log M = 9.0--10.0\,\Msun: these are observed to have rising SFRs  since 
$z=1$, during which time most of their stellar mass was produced.  There is considerable diversity in each mass range; in fact, 
examples of falling, constant, and rising SFHs can be found in every panel.}

\end{figure*}

In this paper we take a next step, presenting what we believe is compelling evidence for SFH diversity, using well-measured
though coarse SFRs and mass build-up for over 22,494 galaxies over the epoch $0.30<z<0.90$ from the Carnegie-Spitzer-IMACS (CSI) 
Survey.  

Following the recent custom in colloquia of presenting conclusions at the start, we move immediately to a graphic presentation 
of our principal result on the diversity of SFHs.  The customary discussion of the data and the analysis that compels this result 
\emph{follow}, and this is used to develop a different, more quantitative description of the data based on the mass of ``old'' 
and ``young'' stellar populations.

\section{Graphic Evidence for SFH Diversity}

The most direct observational test of conformity in galaxy SFHs is to compare their SFRs over a wide range of epochs.  The spectral
energy distributions (SEDs) measured in the CSI Survey span from restframe ultraviolet (UV) to through the near-infrared (NIR) and 
include low-resolution spectrophotometry that provideds exactly this information.  We present in this section a diagram of 240 
randomly selected SFHs from CSI Survey data that represent what we believe is a decisive test of the conformity versus diversity issue.  

The CSI Survey and its methodology are thoroughly described in Kelson \etal\ (2014a, hereinafter K14) and in greater detail in Section 4 
than described now. The subsample of CSI we use here is representative: it is flux-limited from Spitzer-IRAC 3.6$\micron$ imaging over 
$\sim$5.3 square degrees in the SWIRE-XMM-LSS field (see K14).\footnote{As described in K14, by using 3.6$\micron$ flux to cover 
the luminosity peak of common galaxies, the CSI survey is  closer to mass-limited than comparable studies.}  Additional photometry in 
the NIR (J and K bands) came from observations with \emph{NEWFIRM}  on the Mayall Telescope  (Autry \etal\ 2003), and red-optical 
to ultraviolet was obtained by reprocessing \emph{ugriz} data from the CFHT Legacy Survey.  The prism mode with \imacs\ (Dressler 
\etal\ 2011)  on the Magellan-Baade telescope  was used to obtain low-resolution spectra (\emph{R} $\sim20-50$) in the 
range $5000\ang\ < \lambda < 9500$\ang\ for $\sim$43,000 galaxies in this field.  The present study uses a 22,494 galaxy 
subsample of these data covering the redshift interval $0.3<z<0.9$.

In K14, stellar population models were made using models from Maraston (2005).  Subsequently the data have been reprocessed using 
the \emph{Flexible Stellar Population Synthesis} (FSPS) model of Conroy, Gunn, \& White (2009) and Conroy \& Gunn (2010), to take 
advantage of its improved modeling of the TP-AGB.  All stellar mass and SFRs are based on a Chabrier (2003) initial mass function.  

K14 parsed the CSI SFHs into six age bins, with model SEDs for each  of these  summed in generalized maximum likelihood fits to the 
observed prism and broadband fluxes simultaneously. The model SED for each age bin was replicated at four quantized levels of  dust  
attenuation in $A_V\in\{0,0.5,1,2\, mag\}$ for a total of 24 stellar population components  (called \emph{base components} in K14 
nomenclature)  being nonnegatively summed in each fit.  Allowing each component to have variable dust attenuation is a crude 
attempt to allow for variable spatial distribution of dust with respect to 
the underlying distributions of stars.  

Each model, described by a \emph{constant} SFR over each of each of six age bins (or \emph{time intervals}), can be effectively 
reduced to mean SFRs on three timescles---$10^{10}$, $10^9$, and $10^8$ years:

\begin{itemize}
\item{SFR1:  $z=5$ for \Tmax\ Gyr (K14 time interval 1)}  
\item{SFR2:  1 Gyr to 200 Myr before \Tobs\ (intervals 2-5) }
\item{SFR3:  200 Myr to \Tobs\ (interval 6) }
\end{itemize}

\noindent{where \Tobs\ is the age of the universe at the time of observation, and \Tmax\ is the duration of the first epoch of star 
formation (a parameter of the fit), starting at $z=5$, 1.2 Gyr, and potentially lasting until 1 Gyr before \Tobs\ (but usually
stopping well before this.)}.  More details about the fitting for redshift, metallicity, and \Tmax\ can be found in 
Section 3. 

These measurments are largely independent: the oldest population (SFR1) is constrainted by the NIR flux, the intermediate 
age (SFR2) comes from the rest-frame optical  photometry and prism spectrum (the light from main-sequence A stars), and 
the ``present" SFR formation (SFR3, averaged over the last 200 Myr) is measured through reframe UV flux below the Balmer break. 

Using these measures of stellar mass growth over different periods of a galaxy's lifetime, we can now look to see the
extent to which galaxies uniformly (conformally) rose and fell: together, or separately?

\subsection{The SFHs of galaxies vary systematically by stellar mass, but with considerable overlap}

The four panels in Figure 1 are plots of log SFR vs. redshift over four ranges of total stellar mass: from log M = 11.0--12.1\,\Msun\ 
(upper left) to log M = 9.0--10.0\,\Msun\ (lower right).\footnote{The total mass is constrained by a full SED, anchored  by broadband 
photometry.} The red, green, and blue points of Figure 1 represent the measured SFRs for timescales of $10^{10}, 10^9$, 
and $10^8$ years.  The mass bins are well populated, with 1785, 8279, 9062, and 3385 galaxies, respectively.  However, for clarity we 
show only 60 randomly selected galaxies chosen to be evenly spaced in redshift, $\Delta$z=0.01.  SFRs of 0.1 to 1.0\,\MsunYr\ 
are detected with low signal-to-noise ratio (S/N), and SFRs $<$ 0.1\,\MsunYr\ are essentially non-detections.  For display we have 
randomly scattered the non-detections in a log distribution from $-1.0 <$ log SFR $< 0.0$, mixing them with the low S/N detections.   
From repeat measurements for thousands of galaxies in the sample (see /S3.1, /S4.0), we find that the typical errors in the log of 
SFR1, SFR2, and SFR3 are 0.17, 0.27, and 0.22, respectively, much smaller than the trends and dispersion in properties of SFHs exhibited 
in Figure 1. 

Systematic trends of the SFR measurements are apparent.  The SFHs of the most massive galaxies (upper left panel) generally 
decline, from the high SFRs of stellar populations that are observed at ages of $\approx\,$2-6 Gyr, to the SFRs of 1-Gyr-old and 200-Myr-old 
populations.  Not all massive galaxies exhibit steeply declining SFHs:  a substantial fraction show gently declining SFRs over a Hubble 
time, perhaps representing more massive, or more vigorously star-forming, versions of the Milky Way (M101? M83?).  There are, however, 
no \emph{young galaxies} in this mass range, that is, ``late bloomers" whose SFRs rose from the old population to the younger ones.  A move in 
this direction begins in the second mass bin, log M = 10.5--11.0, characteristic of the Milky Way mass today, where the SFRs typically
fall more slowly than for the most massive sample, and there are $\sim$5 out of 60 galaxies  that are ``young.''   The typical SFHs in the 
top right panel look about as we speculated (in Section 1) for the Milky Way: an initial period with SFR $\sim$ 10-20 \MsunYr,
 declining steadily to perhaps $\sim$3 \MsunYr\ at $z\sim0.6$, headed toward today's level of $\sim$1 \MsunYr. 

The bottom two panels, on the left, log M = 10.0--10.5 (a factor of 2-3 below M$^* = 3 \times 10^{10}$ at $z=0.6$---Tomczak \etal\ 2014) 
and on the right, log M = 9.0--10.0, show a large cohort of SFHs that are slowly declining (M33?) or near constant (NGC253?).  But, 
here the ``young galaxy'' population has become obvious: $\sim$50\% of the population in the log M = 10.0--10.5 (lower-left panel)
are '`young,"  and these rising SFHs dominate the lower right panel. 

Because they are remote, some SFRs measured for old stellar populations (SFR1, the red points) suffer a systematic error relating 
to how well the SED fit was able to decouple SFR1 from \Tmax, the timescale over which that mass was produced:the two quantities are 
covariant.   Repeated measurements (discussed in Section 4) indicate that \Tmax\ is able to distinguish between old populations and very
old populations with limited fidelity.\footnote{The \emph{product} SFR1 $\times$ \Tmax, which is the mass produced in this 
epoch, is well measured, as we also show in Section 4.}  Among the data are very high SFR values (exceeding 100\,\MsunYr), the result 
of improbably short \Tmax\ (less than 1 Gyr---an artifact of the fitting procedure), which biased the SFRs high.  Figure 1 would have 
included a small fraction of  these inflated SFRs; the correction we made (by redistributing the \Tmax\ $<1$ Gyr values in the range 
$1< \Tmax < 2.5$ Gyr) lowers only these highest SFRs---there is no effect on the SFHs with \emph{rising} SFRs, our principal 
interest.  In Section 4 we sidestep this covariance problem by switching from SFRs to comparing the ratio of ``old'' and ``young'' stellar mass.  

Figure 1 shows, then, that a substantial population of galaxies in the lower two panels are genuine cases where no ``old'' star formation 
(SFR1) has been detected  (in effect, an upper limit of $\sim$5$\times10^9$\,\Msun\ in ``old'' stellar mass, as we discuss in Section 4).  It is this 
population of ``young galaxies" that O13 inferred and \emph{required}, based on comparisons of sSFR distributions from $0 < z < 1$.  

This, then, is the crux of this paper, to show specific cases---thousands---where in situ growth after $z=1$ dominates the stellar mass.  
These are not simply cases of ``downsizing," that is, extended star formation over a Hubble time, which could be considered scaled SFH 
copies of more massive galaxies on a falling SFMS---a conformal population.  Rather, these are galaxies that peak late, or have not peaked
at all, and for which star formation lasts a few gigayears before fading by the present day.  These are the ``late bloomers" of the G13 
lognormal SFHs---galaxies with long \T0 and short $\tau$---that are a certain sign of diversity in SFHs (see Figure 10, discussed 
in Section 5). 


\begin{figure*}[t]

\centerline{
\includegraphics[width=8.0in, angle=0]{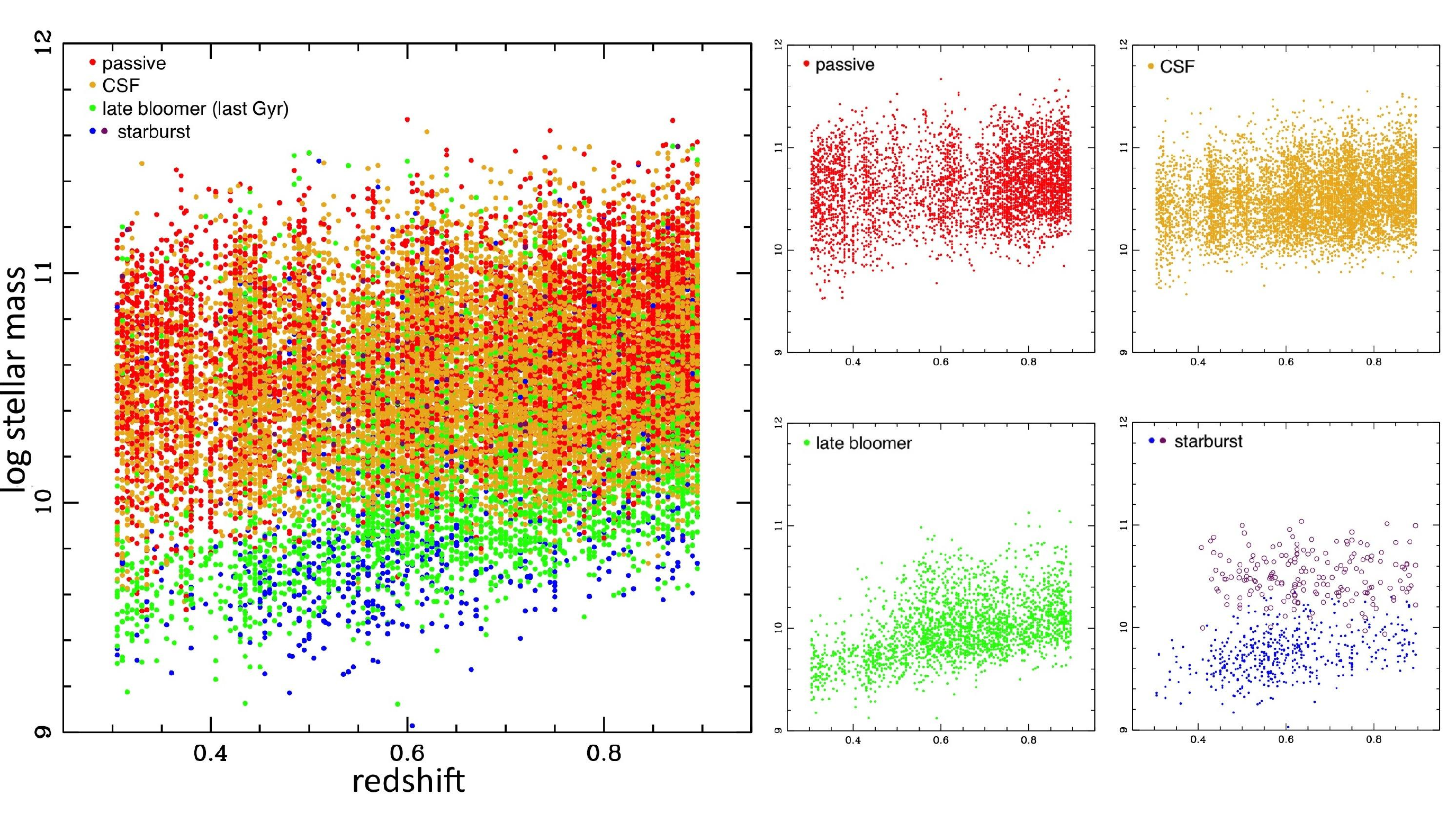}
}

\caption{ Large plot: color-coded SFHs for 22,494 galaxies in the XMM field, 0.3 $<z<$ 0.9.  Four right-side plots; galaxies dominated 
by old stellar populations (top left panel, red points) cover more than an order of magnitude of stellar mass: such ``old galaxies" are found 
with approximately uniform density  over the mass range $10^{9.5}$\,\Msun\ $<  M_* < 10^{11.2}$\,\Msun.   
Slowly declining, or constant SFHs (top right panel, amber points)---galaxies that are still forming stars at \Tobs---also cover a wide mass  range 
that shifts only slowly to lower mass, although the number density of these systems is declining at a moderate rate over the redshift interval 
$0.3<z<0.9$.   Genuinely ``young galaxies" (bottom left panel, green points), whose SFRs peak after $z=1$, are concentrated in lower 
masses---at or below $10^{10}$\,\Msun---a distribution that \emph{is} shifting to lower masses with cosmic time.  Also noteworty are 
young galaxies as massive as today's Milky Way, which rise in prominence to $z\approx0.6$ but are disappearing by $z\sim0.3$.  Identified 
by their very high SFRs in the final 200 Myr epoch,  starburst galaxies (bottom right panel) spread uniformly over a wide mass range for the 
CSF population (open purple circles), and for the young galaxy population (blue points), again over a wide range of (declining) stellar mass .
}

\end{figure*}

\subsection{Four SFHs Made from Three SFRs}

There is also evidence from Figure 1 that---although SFH diversity as a whole shifts to more extended histories with decreasing
mass---there is considerable SFH diversity at \emph{the same stellar mass}.  We see this expansive diversity at similar stellar mass 
as \emph{fundamental}.  In conformal models, the scatter in the SFMS is treated as inconsequential.   

To explore this behavior, we define four SFHs representative of the diversity we see in the graphic tracts in Figure 1.  In what follows, 
\emph{sfr1, sfr2} and \emph{sfr3} are the log$_{10}$ values of SFR1, SFR2, and SFR3.  The categories, for illustrative purposes (not to be taken 
as definitions) are as follows:

\begin{enumerate}

\item{RED = old galaxy (falling SFH)}
\newline{\emph{sfr1} $>$ 1.0 \ and (\emph{sfr1} - \emph{sfr2}) $>$ 1.0}
\item{AMBER = CSF (continuous SFH)}
\newline{(\emph{sfr1} + \emph{sfr2}) $>$ 1.0 and abs(\emph{sfr1} - \emph{sfr2}) $<$ 1.0}
\item{GREEN: young galaxy: (constant or rising SFH)}
\newline{\emph{sfr1} $<$ 0.0 and \emph{sfr2} $>$ 0.5 $||$ \emph{sfr3}  $>$ 0.5}
\item{BLUE = starburst}   (or)  \,    PURPLE = starburst
\newline{AMBER and \emph{sfr3} $>$ 1.5 $||$ GREEN \& \emph{sfr3} $>$ 1.0} 

\end{enumerate}

\noindent{Figure 2 plots our entire CSI sample broken down into these categories.}\footnote{Large-scale structure features, such 
as the broad overdensity at $z\approx0.65$ and the higher-density cluster/supercluster at $z\approx0.37$, are familiar features 
in such diagrams.} 

The trend with mass is unmistakable: from old stellar population dominance, through continuous star formation (CSF) over a Hubble 
time (declining---Milky Way-like, or constant---NGC253, LMC-like?), to genuinely young galaxies.   An added ``frosting" of starburst galaxies 
is found for both the CSF and ``young'' galaxy populations (but rarely for ``old'' galaxies).

To quantify this population of young galaxies, we will in the sections to follow shift the emphasis from SFRs to stellar mass build-up.  
Before proceeding, however, we return to the point made through Figure 1, now abundantly on display in Figure 2, that the diversity 
of SFHs is a clear function of stellar mass.  But, more than this, it is the overlap  of these populations, shown by decomposing the left-hand 
multicolor panel into its four components, that is arresting.  ``Old'' galaxies, those with strongly declining SFRs, cover more than an 
order of magnitude in total stellar mass, limited perhaps only by CSI depths and sensitivities, and CSF span an order of magnitude in 
stellar mass as well.  The most remarkable fact is that young galaxies, although not common, are found up to and beyond the 
mass of the Milky Way.\footnote{Also striking, and provocative, is that these more massive ``young'' galaxies rise in prominence at 
$z\sim0.6$ and fade by the present epoch, clear examples of the long \T0\ and short $\tau$ behavior that can occur in the lognormal 
picture.}  In other words, even galaxies as massive as the Milky Way---not just low-mass galaxies---can follow strikingly different SFHs.   
Our CSI data appear, at a minimum, to confirm the general breakdown of conformity described by Abramson \etal\ (2016) in their 
interpretation of SFMS scatter as a consequence of the lognormal SFH model.

From that conclusion, it is a short step to postulating that galaxies with the same dark-matter-halo mass evolved along very different 
SFHs, raising questions about the fidelity of the popular practice of ``abundance matching" (e.g. Conroy \& Wechsler 2009). If true, 
what drives this diversity?  We are drawn to the notion that such differences are built in from an early epoch (as promoted by Dressler 
in his 1980 study of the morphology--density relation), perhaps related to the density of the dark-matter halo or larger-scale properties 
of the spectrum of density perturbations in a galaxy's neighborhood.  Although this possibility is not compelled by the results of
this study, we will explore some aspects of this problem later in the paper.  

To this point we have examined SFHs in terms of SFRs in three distinct epochs, acknowledging that a star-formation history is 
the evolution of the star-formation rate.  However, a more quanitative evaluation using the mass fractions of old and young stars 
will serve better in understanding what these histories mean in terms of building galaxies and in the comparison to models.  We 
invite the reader to continue to appreciate the full implications of diversity in SFHs.

Section 3 describes in detail the CSI data and how they are analyzed to provide (independent) SFRs over timescales of
$10^{10}, 10^9$, and $10^8$ years.  In Section 4 we turn the analysis to the build-up of stellar mass, showing how the ratio of two 
integrated masses---the stellar mass produced from $z=5$ to 1 Gyr before observation, and stellar mass produced in the last 
gigayear---provides a very good tool for searching for the ``late bloomers" that  O13 `required.'  While O13 was unable to  point to 
specific galaxies that had rising SFRs after $z=1$, we \emph{can} identify individual galaxies of this population.  In Section 5 we look for  correlations of  SFHs related to environment, specifically, signs of a centrals/satellites dichotomy.  In Section 6 we examine the consistency 
of the G13 model with CSI SFHs in terms of the full range of SFHs we find.  Finally, in Section 7 we summarize our results on SFH diversity 
and review the implications for conformal (plus quenching) models.

\section{The Data: SFH\MakeLowercase{s} from the Carnegie-Spitzer-IMACS Survey}

The defining characteristics of the CSI Survey were large areal coverage   (three widely separated \emph{SWIRE} 
fields of 5 sq deg each (see K14), galaxy selection via Spitzer-IRAC 3.6$\micron$ flux, and near-IR to optical-ultraviolet 
photometry and low-resolution  spectroscopy from 4500\,\AA\ to 10,000\,\AA, corresponding to rest frame 3462\,\AA\ at 
the $z=0.3$ lower limit and 5263\,\AA\ at the $z=0.9$ upper limit, for the subsample used in this paper. Flux measurements at 
3.6$\micron$ sample the peak of the energy distribution in SEDs for typical galaxies at $z<1$.  Because of this, the stellar mass 
completeness limit changes much more slowly with redshift than in previous spectroscopic surveys selected by \emph{I}-band 
or \emph{R}-band magnitude, as shown in K14.  Over the $0.3<z<0.9$ interval of our subsample, this limit changes only by a 
factor-of-two, from 1 to 2 $\times 10^{10}$\,\Msun.

Figure 3 of K14 shows that a galaxy of $4\times10^{10}$\,\Msun---the present-day mass of the Milky Way---can be detected to $z=1.5$ in 
CSI, compared to $z\ls1$ for PRIMUS and DEEP2.  It is this combination of depth and volume, 0.14 Gpc$^3$ for the full survey (an 
order of magnitude more than DEEP2), that makes CSI a powerful tool for galaxy evolution studies.

The CSI Survey, through its connection to \emph{SWIRE}, is a \emph{field} survey, that is, it does not include regions of high galaxy 
density, for example, a rich cluster.  There is a substantial population of groups, however, that have been characterized 
by Williams \etal\ (2012) and Patel \etal\ (2016).


\begin{figure*}[t]

\centerline{
\includegraphics[width=6.0in, angle=0]{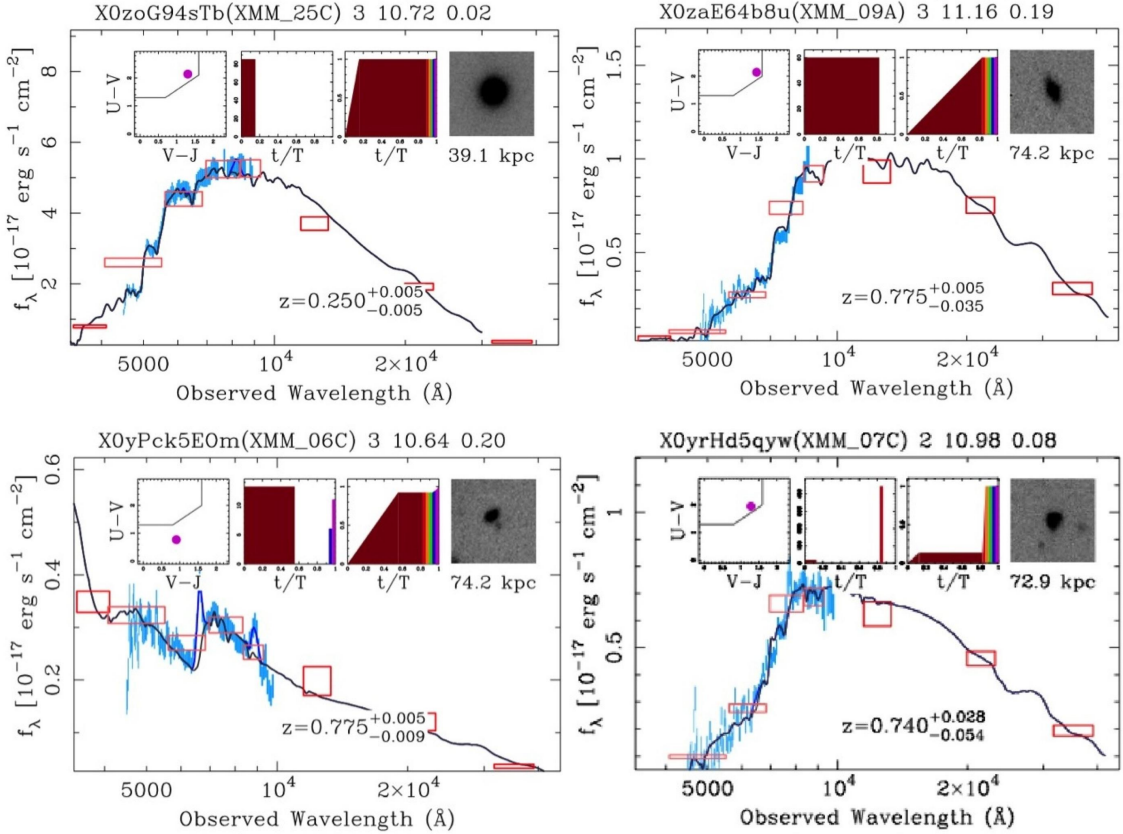}
}

\caption{Examples of data and SEDs from the CSI survey.  Each panel shows a plot of $f_\lambda$ versus wavelength for 
photometric data (red boxes), spectroscopic data (cyan trace), and model SED (black line).  The four small boxes above these 
contain derived data: (left) the UVJ diagram that separates star-forming from passive galaxies; (middle left) constant SFRs derived 
by the model for six time intervals, as described in the text; (middle right) integrated SFRs (stellar mass build-up) of the 
model; (right) the galaxy's \emph{z}-band image from the CFHT Legacy Survey.}

\end{figure*}

As with many astrophysics programs these days, the CSI Survey is considerably oversized compared to the complexity of the phenomena 
studied here.  For our work on SFHs, it is the quality of the photometric and spectroscopic data---and how they enable robust 
measurements of stellar mass---rather than the sample size\footnote{The full SWIRE-XMM-LSS catalog of the first data release of 43,347 
galaxies covers $0.3<z<1.5$.} or areal coverage, that are the strengths of the CSI Survey.  Our subset of the CSI sample of galaxies studied 
in this field reduces to 22,494 galaxies over the redshift range $0.3<z<0.9$. Of  course, the rarest galaxies, with stellar masses of 
M $>10^{11}$\,\Msun, and those with M $<5\times10^{9}$\,\Msun---approximately the 50\% completeness limit of the Survey---are 
represented by only 1728 and 294  galaxies, respectively, but for the most important middle territory of this work, the sample size does 
not limit our ability to address the primary question of SFH diversity.

Figure 3 shows observational data and model SEDs for four representative galaxies in the survey, including the six discrete intervals 
of star formation (the \emph{base components} for fitting stellar population models  to the SED, described below) that are used 
to measure SFRs over timescales of $10^{10}$, $10^9$, and $10^8$ years.  Each box is a plot of observed wavelength and 
$f_{\lambda}$ flux.  The observational data in each box are the eight bands of photometry (red boxes) and the \emph{IMACS} 
prism spectrum (cyan trace).  

A large part of the data-reduction challenge went into intelligently combining (slit) spectral data and aperture photometry data and 
validating the integrity of the combined \emph{spectrophotometric} data, as described in detail in Section 3 of K14.  The key step 
was to use the aperture photometry to anchor and correct imperfect flux calibrations of the prism spectra (affected by slit losses and 
differential refraction).  The accurately fluxed prism spectra enable SED fits to resolved spectral features that constrain the model 
of stellar populations far more tightly than possible with photometry only, even with a large number of photometric bands.

As described briefly in Section 2, K14 made generalized maximum likelihood fits\footnote{The generalization was in using Huber's 
M estimator instead of L2, the standard $\chi^2$ M estimator used in most likelihood analyses.} of these SEDs, starting with Maraston (2005) 
stellar libraries (K14, Section 4).\footnote{Subsequent to K14, these libraries have been replaced with those constructed by Conroy \etal\ 
(2009) and Conroy and Gunn (2010).}   An SED model to fit these data was made in a three-dimensional grid of redshift, metallicity,  
and (log of) \Tmax---the duration of the first epoch of star formation---that we have discussed above.  The critical parameters in the model 
were six discrete intervals of constant star formation, starting at $z=5$ for and lasting for \Tmax\ Gyr, and for five 200 Myr intervals 
beginning 1 Gyr before \Tobs.  Each SFR component was replicated with four levels of dust extinction,  with nonnegative sums of these 
able to reproduce the variable dust attenuation of typical galaxies; these are the 24 base components that are fit in the grid 
of redshift, metallicity, and \Tmax.   Figure 6 in K14 is a diagram in cartoon form with examples of SFRs in the six intervals that 
characterize  normal star-forming galaxies, quiescent galaxies, and starburst and poststarburst galaxies.

The four galaxies in Figure 3 also show examples of the modeling process.  The middle two of the four small boxes at the top of  each 
Figure 3 panel show the star-formation and stellar mass growth histories for that galaxy through the SED fitting procedure.  The 
far left box shows the position of the galaxy in a UVJ diagram (Williams \etal\ 2009).   Galaxies in the upper left region are considered 
passive, while outside it they are star-forming.  The far right box shows a \emph{z}-band image of the galaxy from the CFHT Legacy Survey.

It is apparent for these four cases that each derived SED is an excellent match to the broadband photometry and blue-to-red spectra.  
K14 emphasizes that the fine detail contained in the spectra plays a particularly important  role in constraining redshift to $\ls$2\%, a better 
performance than is possible with broadband photometric data alone.  Redshifts of this accuracy are essential for sampling the prism 
spectra at the proper wavelengths, and vice versa.  This degree of accuracy in the redshift was also intended as an important feature of the 
CSI Survey, because it allowed the association of galaxies into groups and clusters.  This environmental component of the survey, which
we exploit in Section 5, is something that cannot be achieved with redshifts of lower precision.

As was remarked in Section 2, the power of the CSI data set with respect to SFHs is bound up in the several stellar evolution timescales
inherent in the data.  The K14 analysis recognized this explicitly by adopting six time intervals, over which the SFR was assumed  to be 
constant, as the \emph{base components} for \emph{solving} the SED.  Five of the six are fixed time intervals of 200 Myr that started 1 Gyr 
before \Tobs: star formation over this period produces strong variations in the blue-to-visible SED that provide very effective constraints on 
the model.  The SFR1 interval, starting at $z=5$, takes advantage of (and suffers from) the immobility of the giant branch in 
color-magnitude diagrams for stellar populations with ages ranging from 2 Gyr to a Hubble time.  Also, the stellar mass  producing 
the light from old populations---the product SFR1$\times$\Tmax---is reasonably well measured (see Figure 6); the degeneracy of the 
giant branch as a  function of age makes \Tmax\ alone difficult to constrain. 

Because of the uncertainty in determining \Tmax, the K14 analysis allowed the (constant) star formation starting at $z=5$ to continue 
up to 1 Gyr prior to \Tobs.  However, the constraint on \Tmax\ improves if light from main-sequence stars makes a significant contribution 
to the SED.  From the vantage point of present-day galaxies, this is never the case, but from a galaxy with a look-back time of 5 Gyr---the 
median for our sample---the contribution from dwarf  F stars (with main-sequence liftetimes of $\sim$5 Gyr) can be detected as far 
back as $z\sim$\,2, the peak epoch of cosmic star formation.  For this reason the CSI spectral data have some leverage on \emph{when} 
the stars of the first interval were formed,  not just how many. \Tmax\ may not be well constrained, but neither is it unconstrained.  
We believe that this is why most \Tmax\ values fall in the reasonable range of 2--6 Gyr, timed to the F-star contribution detected in the 
blue-to-visible part of the prism spectra.  This would also explain why a there is a small  but significant population with \Tmax\ $<1$ Gyr; 
such dubious measurements are likely to be the default ``$\delta$ function'' at $z=5$ when there is no detectable signal from 
an old F-star population (exacerbated, we think, by the discretization of \Tmax\ into small logarithmic intervals as log \Tmax\ approaches 
zero).


\begin{figure*}[t]

\centerline{
\includegraphics[width=4.0in, angle=0]{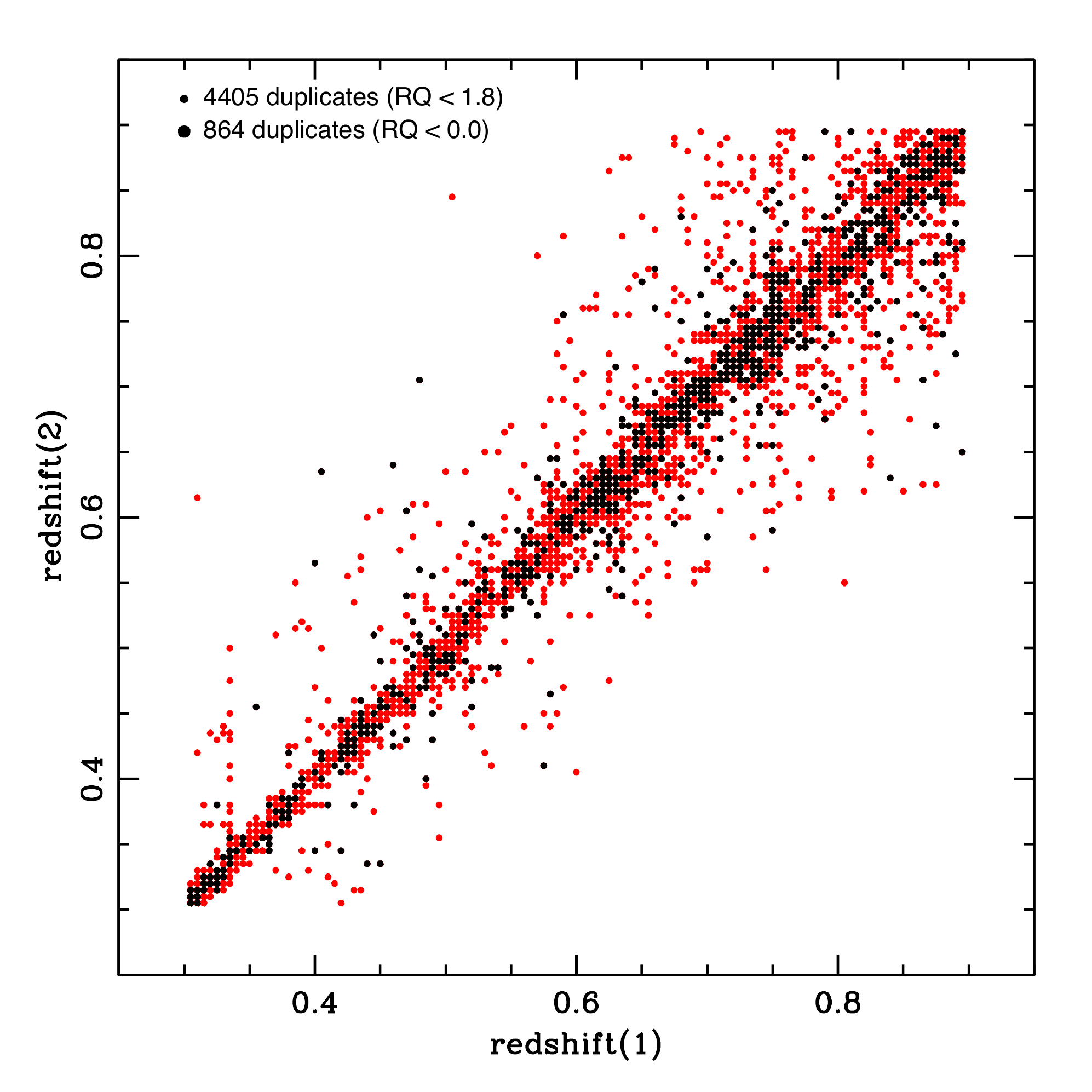}
}

\caption{Comparison of redshifts for  duplicate measurements ($R < 2.5$, red points) and 864 measurements ($R < 0.0$, black points).  
Points are overplotted because of quantization of redshifts at 0.005.  The scatter of 2.2\% at the characteristic redshift of the sample, 
$z\sim$0.65, is in good agreement with that found in K14.
} 

\end{figure*}

Examples can be seen in Figure 3 in the two middle boxes above each SED. The left-hand box  shows the SFH expressed as 
constant SFRs in six time intervales, as explained above, and the right-hand box shows the  integral of the SFH, for example, the build-up of 
stellar mass.  The galaxies of the top two panels contain only old stars, according to the model.  The case on the left is one of
prompt star formation in the first few gigayears (short \Tmax), while for the galaxy on the right the model has settled on a long history
of early star formation that continues up to the last Gyr before \Tobs.  No star formation has been detected in any of the final five 200 Myr
time intervals.  (The integrated mass does not increase in the five colored bins.) Both galaxies fall into the ``passive" region of the UVJ 
diagram (Williams \etal\ 2009).  It is not obvious looking at the prism spectra why the SED fiting process has extended SFR1 star formation 
to relatively late times.  This could be a sign of the F-star contribution just discussed, but this effect is expected to be subtle.  We are 
working on a variation of the D4000 index suitable for such low-resolution data in an effort to tease out the subtle markers of star formation
that occurred a 1-3 Gyr before \Tobs.

In contrast, the galaxies in the two bottom panels appear to have had considerable late-epoch star formation.  The galaxy on the left 
has accumulated most of its $\sim$4 $\times 10^{10}$ \Msun\ from between 1 and 4 Gyr after the Big Bang according to the model fit.  
The star formation in the last two intervals, starting 400 Myr before \Tobs, add only a little mass (see the right box), but the changes to the 
SED are dramatic.  In particular, the strong upturn in the ultraviolet flux and the peaks in the SED from [O\,II] and [O\,III] emission are 
indicative of vigorous star formation in the interval that records the final $2\times10^8$ years.  The galaxy is firmly in the ``star-forming''
part of the UVJ diagram.  

The galaxy on the bottom right is a good example of the principal result of this paper: this massive galaxy of 10$^{11}$ \Msun\ is
almost entirely composed of young stars, according to the model fit, which records only a small amount of star formation up to 1 Gyr
\emph{before} \Tobs, with all the rest in the last gigayear.  Note also the lack of star formation in the final (200 Myr) interval, as the absence of 
an ultraviolet upturn confirms that this galaxy is in the ``passive region'' of the UVJ diagram.  Although the SED looks similar to the passive 
galaxy directly above it, a closer look at the blue part of the spectrum reveals important differences: the D4000 break and G band, 
so prominent in the upper right example of an old galaxy, are not present in the bottom right galaxy.  The spectrum of this galaxy is 
essentially pure A stars, the unambiguous signature of a $\ls$1 Gyr population.

\subsection{Repeated Measurements}

The data just discussed were drawn from a catalog of 33,089 measurements of objects at 0.3$<z<$0.9 in the XMM field.
This included 4665 repeated observations and 366 galaxies with three measurements.   From the extensive error analysis 
described in K14, a quality parameter \emph{R}(S/N) was used to pare the sample to $R < 1.8$ (slightly better than 
the $q=2$ rating in a three-tiered system---see K14, Figure 8) for single observations.  Multiple observations were combined 
when $R < 2.5$ (slightly worse than $q=2$ for each measurement).  

Our final catalog contains 22,494 individual galaxies, where the 27,692 unique objects have been further trimmed by 5198 
galaxies below the quality cut.


\begin{figure*}[t]

\centerline{
\includegraphics[width=5.0in, angle=0]{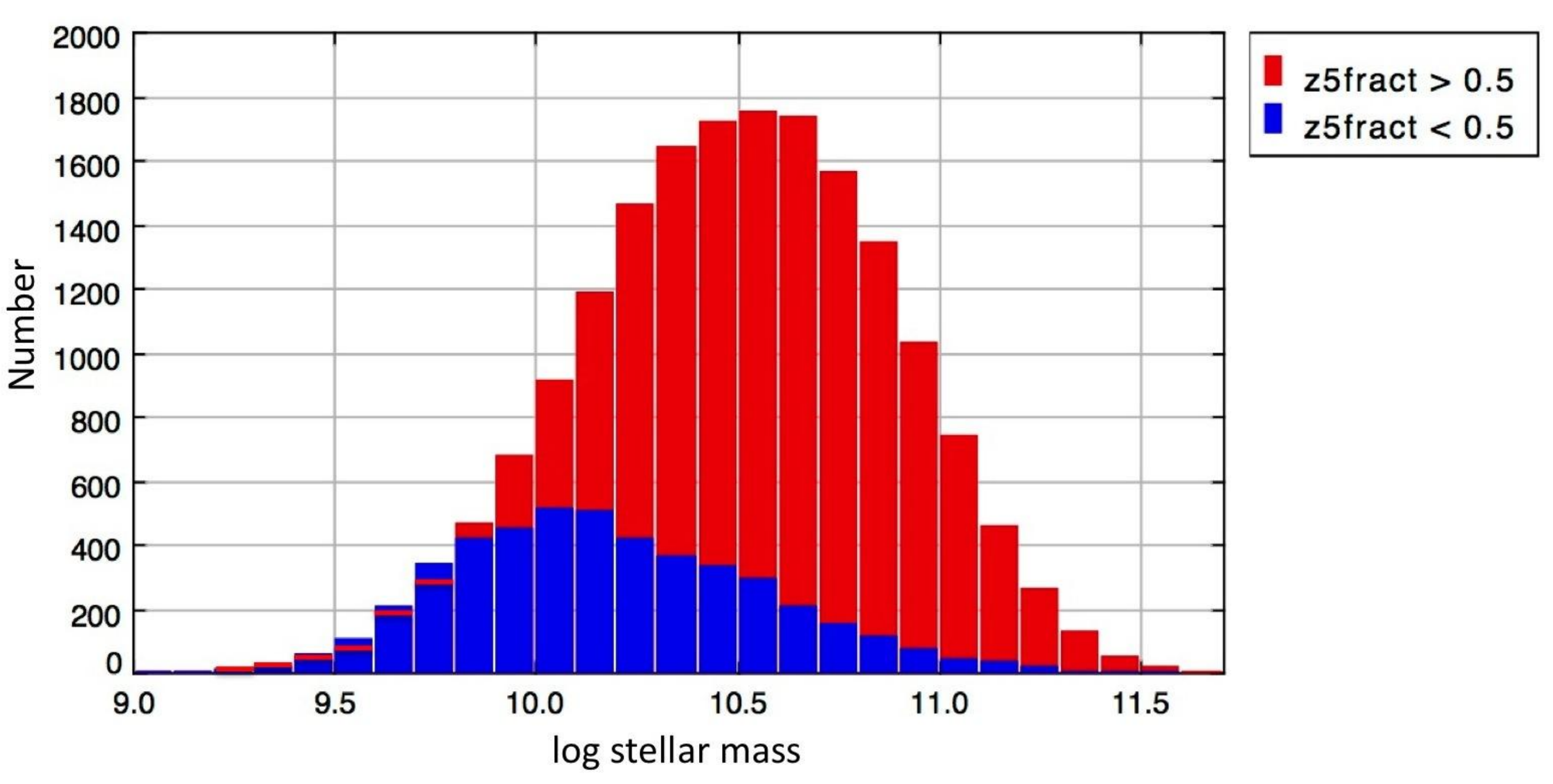}
}

\caption{Distribution in mass of the ratio of old to young stellar populations.  The galaxies in red have (at \Tobs) most of their mass in 
old stars, \emph{z5fract}$>$0.5, while the galaxies in blue are dominated by stellar mass produced in the final gigayear.  As found for the
SFHs in Section 2, the younger galaxies are systematically lower in mass, but, as also found in Section 2, the overlap in mass is large and real.  The
factor of two or better measurements of old and young stellar mass, over appropriate ranges, make this a robust measurement, as 
explained in the text.}

\end{figure*}


\begin{figure*}[t]

\centerline{
\includegraphics[width=5.0in, angle=0]{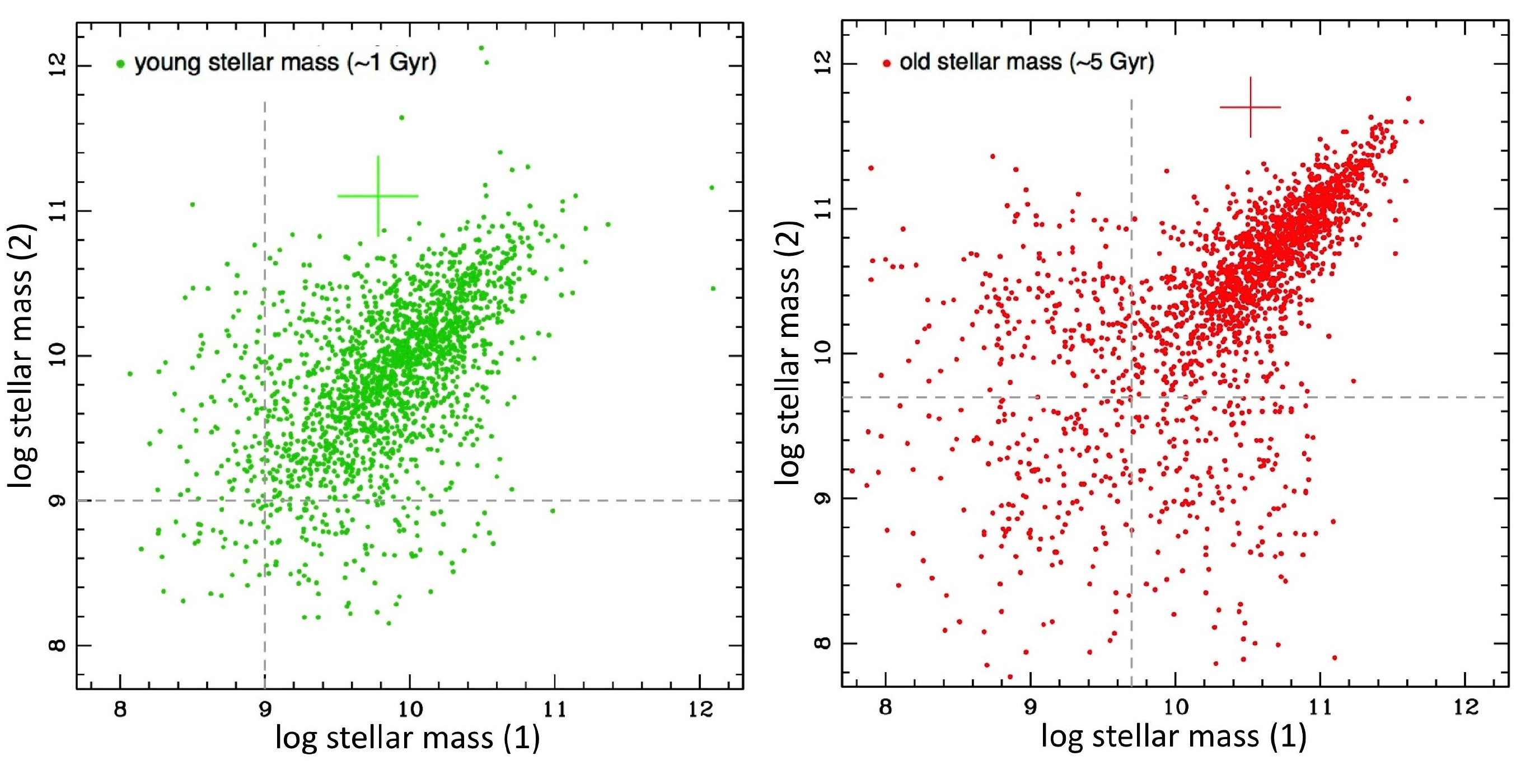}
}

\caption{Duplicate measurements of stellar mass for two epochs. Right:  $z=5$ (\emph{T} $\approx1.2$ Gyr) to (as late as) 1 Gyr 
before \Tobs. Left: The final gigayear before \Tobs.  The crosses represent $\pm1\sigma$ errors, over the mass range $M_*>10^9$\,\Msun\
for the ``young'' stellar mass and $M_*>10^{10}$\,\Msun\ for the `old stellar mass.'  These mass limits are shown by the dashed gray lines;
the slightly lower $M_*>5\times10^9$\,\Msun\ limit shown in the right-hand diagram marks the point below which the mass measurements
begin to strongly diverge.  It is unsurprising that the detection limit for an old stellar population is higher, since most of this mass is viewed 
from an epoch $\sim$5 Gyr from \Tobs.  Typical errors in these measurements, 0.28 in the log (a factor of 1.9) for the young (green) 
population, and [0.21] (a factor of 1.6) for the old (red) population, are sufficiently good to constrain the \emph{zfract} parameter (the 
fraction of stellar mass produced early compared to the total) to a factor of 1.5--2.0 over its range.}

\end{figure*}

Figure 4 plots the redshift determined for each of the 4666 galaxies with duplicate measurements. From the statistics of the differences 
in the pairs of measurements, we find $\sigma=0.022$ for the redshift error per observation. For the median redshift of the sample 
$z\approx0.65$ this value is in good agreement with the $\sigma/(1+z) = 0.011$  found in the extensive error analysis done by K14.

\section{The Build-up of Stellar Mass Since \lowercase{z}$=5$}

As described in Section 2, the SFR in the first $\sim$5 Gyr is covariant with \Tmax, the SED model's fit to the duration of star formation that 
began at $z=5$,  $T \approx$ 1.2 Gyr.  On the other hand, the \emph{stellar mass} in this and later epochs is well constrained by the infrared portion of the SED.  To exploit  this and to quantify the diversity of SFHs, we chose to characterize the galaxies simply in terms of ``old'' 
and ``young'' stellar mass, defining the first  as star formation that began at $z=5$ and lasted typically 2--6 Gyr for this sample, and 
the second as star formation taking place in the last gigayear before \Tobs.  We then define \emph{z5fract}, the integral of the stellar mass 
produced as


\begin{equation}
    z5fract \equiv \frac{Tmax\,{\rm SFR1}}{Tmax\,{\rm SFR1} + (0.8\,{\rm SFR2} + 0.2\,{\rm SFR3})}
\end{equation}
where all SFRs are in Gyr$^{-1}$.  That is, {\it z5fract} is the ratio of old/(old+young) stellar mass.
  
Figure 5 shows this result through a histogram of the mass distributions of galaxies with more than 50\% (red) and less than 50\% (blue) 
of their mass in an old population.  There is a clear mass dependence, analogous to that found for the SFHs in Section 2, in the sense that the 
fraction of ``young'' galaxies rises to lower mass: at approximately 10$^{10}$\,\Msun\ there is a 50/50 mix of  galaxies that formed 
most of their mass early and those that formed it late.  This simple parameterization confirms and quantifies the results based on 
SFRs we presented in Section 2: (1) there is a diversity of SFHs at all galaxy masses above 10$^9$\,\Msun; and (2) a significant fraction of 
galaxies at 0.3$<z<$0.9 are \emph{young}, that is, they formed most of their stellar mass after $z=1$.

With the 4666 repeated measurements in the CSI \emph{XMM} sample, as restricted to $0.3<z<0.9$, it is straightforward to determine how
well these two epochs of stellar mass are constrained by the SED fitting process.  Figure 6 plots the derived stellar mass for pairs of measurements 
of the old and young populations.  Above 10$^9$\,\Msun, the stellar mass within the final gigayear (left) shows a scatter of 0.28 dex, or a factor 
of 1.9; the increase in scatter at the low-mass end is dominated by the high-redshift part of the sample.  The scatter for the old population 
above 10$^{10}$\,\Msun\ shown in the right plot is only  0.21 dex, or a factor of 1.6.  The uncertainty increases rapidly below 
10$^{10}$\,\Msun; not surprisingly, it is very difficult to detect $\sim$10$^9$\,\Msun\ of old stars in an intermediate-redshift galaxy.   
These expected errors are appropriate uncertainties for the $\sim$17,000 individual measurements of the 22,494 galaxies in the sample, 
with expected improvement for the $\sim$5000 multiple measurements that we have coadded. 

Based on the factor of two or less uncertainty in the measured masses, the result in Figure 5 is robust.  A possible concern is the fact that, 
in the high-redshift end of our sample, the CSI survey has significant incompleteness for ``passive" galaxies with masses $\sim$3 $\times$ 
10$^9$\,\Msun.  The worry might be that the ``young galaxy'' phenomenon is exaggerated by a failure to detect an old population in these 
low-mass galaxies.  However, Figure 5 shows that the majority of such galaxies found in our study have total masses of 10$^{10}$\,\Msun\ or 
greater; over most of this redshift range, even the ``old'' mass is reliably detected.  For the galaxies with total mass 10$^{10}$\,\Msun, 
the $<$1 Gyr population is secure, which means that, as \emph{z5fract} trends lower, the fraction of mass left for the ``old'' population 
is small: underestimates of that mass have no consequence, and overestimates have a small range because masses above 5 $\times 10^9$\,\Msun\ 
should have been detected.  The potentially missing galaxies are the ones that have total masses below 5 $\times 10^9$\,\Msun\ with little 
or no younger population.  

In summary, while there might be more ``old galaxies" missing through incompleteness, the ones below 10$^{10}$\,\Msun\ that 
we identify as `young' (\emph{z5fract} $<$ 0.5) are secure.

Lower limits of 10$^9$\,\Msun\ and 10$^{10}$\,\Msun\ for masses of the young and old stellar populations are sufficient to determine the 
fraction of old stellar mass for our sample.\footnote{This is the stellar mass at  intermediate redshift, expected to grow modestly by the 
present epoch.}  By comparing 4666 pairs of \emph{z5fract} values, we find that these have typical  errors of a factor of 1.5--2.0, 
accurate enough to reliably separate the  sample over over its full range.  


\begin{figure*}[t]


\centerline{
\includegraphics[width=6.5in, angle=0]{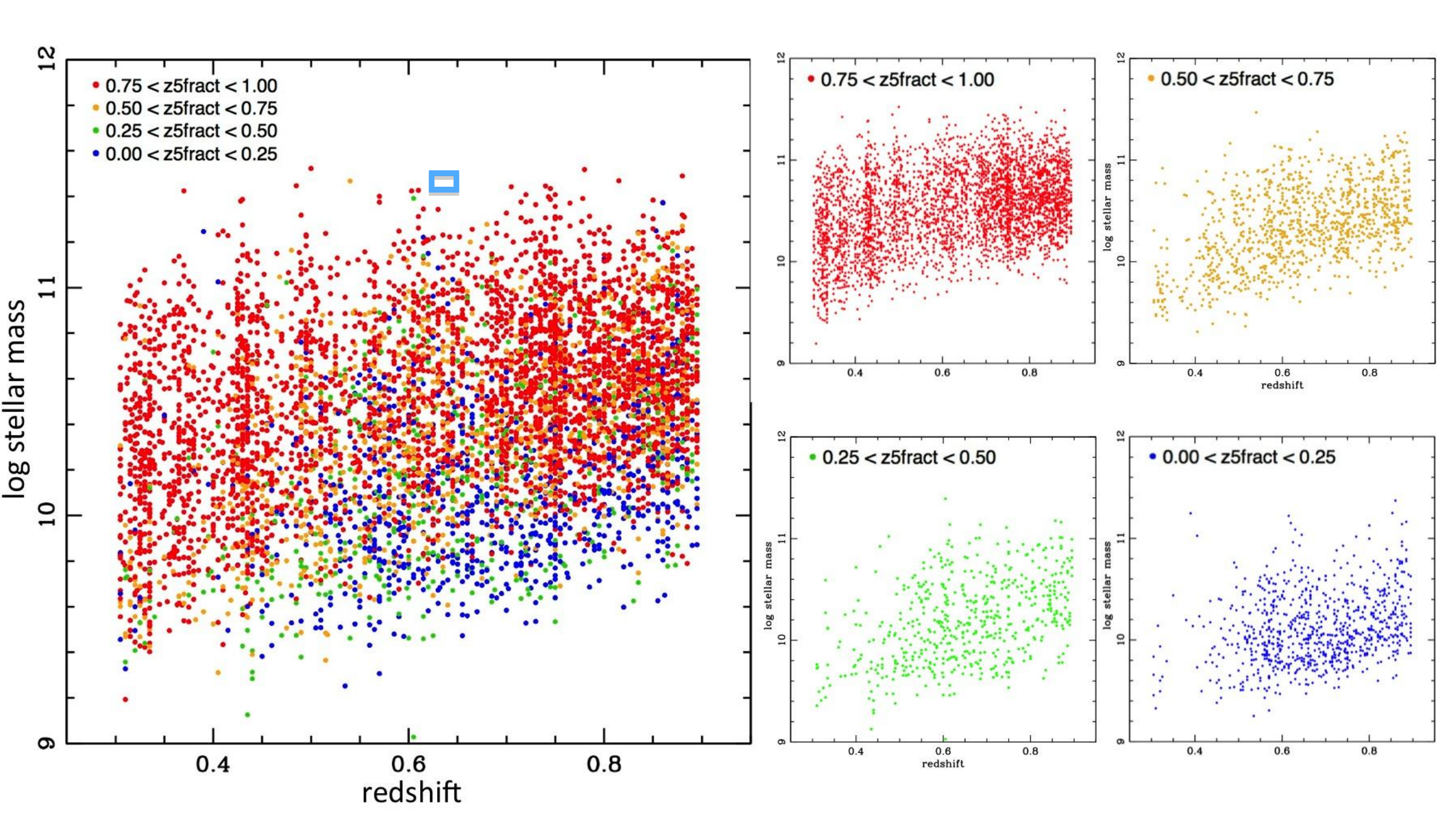}
}

\caption{Large plot: SFHs represented by the fraction of mass produced early, from $z=5$ for \Tmax\ Gyr, divided by the total mass, the 
old stellar mass added to that formed in the last gigayes before \Tobs.  A random sample of 6000 of the 22,493 galaxy sample is shown.  
Right-hand plots: the 0.75 $<$ \emph{z5fract} $\leqslant$ 1.00 population (upper right panel, red points) dominated by old stellar mass 
is found for more than an order of magnitude of the mass range, as in Figure 2. The falling trend with redshift for the highest masses 
for this population is mostly the result of a smaller sample volume with declining redshift that fails to contain these rarest galaxies, while for 
the lower-mass galaxies it represents mass-incompleteness for the more distant galaxies.   The steeper fall in mass of the distribution of 
the 0.50 $<$ \emph{z5fract} $\leqslant$ 0.75 population (upper right panel, amber points) is identified with the continuous-star-forming 
galaxies in Figure 2.  A likely explanation for the steeper trend of this distribution relative to the Figure 2 counterpart (and relative to 
the other distributions seen in this plot) is the transition of many of these galaxies to the (shifted) passive population over time.  This 
is consistent with the shallower dependence for the 0.25 $<$ \emph{z5fract} $\leqslant$ 0.50 and 0.0 $<$ \emph{z5fract} $\leqslant$ 0.25
populations (green and blue  points in the two bottom panels).  SFRs in these younger galaxies fall into two categories: (1) falling slowly 
because of a late peak in SFR ($z\sim1$, i.e., a large \T0) and long histories of star formation (i.e., large $\tau$)---as seen in the blue panel; 
or (2) galaxies that also peak late, but with short histories (i.e., small $\tau$)---as seen in the green panel.  A striking indication of this 
latter population is the peak in numbers of massive galaxies at $z\sim0.6$ in the 0.25 $<$ \emph{z5fract} $\leqslant$ 0.50 panel, a 
population that seems to disappear by $z\sim0$.  These are the young galaxies postulated by O13, modeled by G13, that motivated 
this study.
}

\end{figure*}

\subsection{A Stellar Age Map of the $z \sim0.7$ Universe}

In Section 2 we went from modest numbers of example SFHs in Figure 1 to a redshift-mass plot (Figure 2) that displayed the entire sample
SFHs as discreet classes, by colors.  Here we do the same, by using points of four colors to display the full range of \emph{z5fract} as a proxy for 
SFHs, over its full range, 0.0 $<$ \emph{z5fract} $<$ 1.0.  

Figure 7 shows four equal ranges that resemble the SFH categories of Figure 2.    While the SFHs in Section 2 were defined by 
ad hoc changes in SFR from one epoch to another, the \emph{z5fract} values coded into Figure 7  record a continuous variation 
in SFH from galaxies that are exclusively ``old'' to those that are largely, if not completely, ``young.'' 

As in Figure 2, the most striking thing about the distribution of the red points, $0.75 < \emph{z5fract} \leqslant 1.00$ (upper left panel) is 
the more-than-an-order-of-magnitude range of mass of this population dominated by old stellar mass.  The ``oldest'' galaxies in our
sample are not limited to the most massive ones; indeed, their frequency of occurrence is roughly mass-independent down to 
a factor of 30 smaller than the most massive galaxies in the sample.

The falling trend with redshift for the highest masses is due to the decreasing volume of the survey, while the bottom envelope 
is indicative of the mass incompleteness of the CSI sample as a function of redshift,  as seen in the modeling of Section 6.

The $0.75 < \emph{z5fract} \leqslant 1.00$ interval includes \emph{all traditional SFHs}, that is, those that were either constant 
or declining with cosmic time (discussed further in Section 6).  For the median redshift of the sample, $z\approx0.65$, the time from 
redshift $z=5$ to 1 Gyr before \Tobs\ is 6.6 Gyr, so a constant SFR of any amount will produce $\emph{z5fract}\approx0.87$.  Declining 
SFHs, such as the inimitable $\tau$ model, will be larger, up to $\emph{z5fract} = 1.0$ for a ``passive" galaxy.  This means that, even 
in this category of the ``oldest'' galaxies, there must be SFHs that had rising SFRs for some time after $z=5$ (those with 
$0.75 < \emph{z5fract} \leqslant 0.87$, neglecting the intrinsic error in \emph{z5fract}).

The same can be said for \emph{all} the galaxies in the well-populated $0.50 < \emph{z5fract} \leqslant 0.75$ category (upper right 
panel): these may have declining SFHs at \Tobs, but in order to reach 25\% to 50\% fractions of young stars in the last gigayear before \Tobs, 
the old population must also have been rising for as long as several gigayears after $z=5$.  This in itself (exclusive of our case for younger 
galaxies) points to a diversity of star-formation rise times for galaxies that are dominated by old stellar mass, and it is inconsistent with
the idea of conformal SFHs that all began with the same growth pattern, long \emph{before} the peak of star formation.  On the
other hand, it is natural in the lognormal SFH model, as we discuss in Section 6.\footnote{It is also expected in models where galaxies 
experience stochastic mass growth (Kelson 2014b).}

It is notable that this \emph{amber} distribution falls more steeply than the distributions in the other three panels.  As we surmise in
Section 6, the SFRs for these galaxies peak in the redshift range $1.0 < z < 1.5$, so over the $0.3 < z < 0.9$ epoch of our sample, the
total stellar mass (at \Tobs) of galaxies that land in this \emph{z5fract} range is falling, as the more massive galaxies transition to
the 0.75 $<$ \emph{z5fract} $<$ 1.0 population.  Once again, however, at every epoch the mass range of such SFHs is wide: the 
points represent galaxies with a substantial range in decline rate.

In contrast, the galaxies with $0.25 < z5fract \leqslant 0.50$ (bottom left panel, green points) and $0.0 < z5fract \leqslant 0.25$ 
(bottom right panel, blue points) do not show the relatively steep trend line seen for the \emph{amber} galaxies.  The shallow 
decline of these distributions with mass seems to align with the notion that many of these are galaxies with a longer \T0\ that are 
observed closer to the peak SFRs.  However, the most  interesting feature of these distributions may be the swell in numbers 
around $z\sim0.6$, which are sensibly identified as the long \T0, short $\tau$ population postulated by O13 and whose identification 
was the main motivation for this analysis of the CSI data.  It is perhaps the most striking feature of Figure 7 that so many of these 
genuinely young galaxies peak at $z\sim0.6$ and rapidly disappear thereafter, to be essentially gone by $z\sim0.0$, at least 
those with M$_*>10^9$\,\Msun.   

Figure 7 is rich in information with respect to SFHs, but a framework is needed to understand the longitudinal nature of the samples
in these diagrams.  As we show in Section 6, the G13 lognormal SFH provides one such tool.

\section{Centrals and Satellites?  Diversity in SFHs With Environment}

The influence of local environment on galaxy evolution remains an important yet unresolved issue.  The morphology--density relation
was interpreted by Dressler (1980) as the result of a very early environment (to explain the very weak trend of population gradients for 
such a wide range of present-epoch local density) that manifests in the $0<z<1$ environment, but is not causally related to it.\footnote{
Ironically, this paper is commonly cited as a demonstrating that local environment \emph{is} responsible for the diversity in galaxy 
morphology and its trends.}  It is of course likely that morphology and SFH are closely coupled, so the earlier work may well be germane 
to the present study. 

As has been pointed out widely in recent years, morphology correlates most strongly with galaxy mass (Dressler 2007; van den 
Bosch \etal\ 2008), which suggests that what has been attributed to the local environment can be largely reduced to a statement like ``denser 
environments host galaxies of higher mass than lower-density regions.''   While Dressler inferred from this that the conditions for making 
more massive, and therefore early type, galaxies were imprinted at a very early epoch, others have focused on specific processes associated 
with dark-matter halo growth.  Van den Bosch \etal\ (2003, 2008; see also Conroy \& Wechsler 2009) was among the first to develop a 
now popular picture in which the capture of lower-mass ``satellite" galaxies by higher-mass ``central galaxies" (and the stripping of the 
halos of the former) is a primary mechanism for shaping galaxy SFHs and, presumably, their morphologies.

While a thorough investigation of the ``centrals and satellites'' paradigm is beyond the scope of this paper, the CSI Survey's 
ability to assess the environments of sample galaxies through its spectrophotometric redshift precision of $\ls$2\%, makes
it straightforward to take a first look at how the SFHs derived in this study fit the picture.   Galaxy number densities
and local mass densities for the 22,494 galaxies of our sample were calculated using projected comoving circles of 1, 2, 
and 4 Mpc.  Figure 8 refigures the relationships of Figure 7 for samples that include only the $\sim$11\% of galaxies that
reside in the densest environments, corresponding to projected densities of  `m2' $>$ 12 $\times 10^{10}$\,\Msun\ Mpc$^{-2}$, 
and `n2' $>$ 2.7 galaxies Mpc$^{-2}$.


\begin{figure*}[t]

\centerline{
\includegraphics[width=6.5in, angle=0]{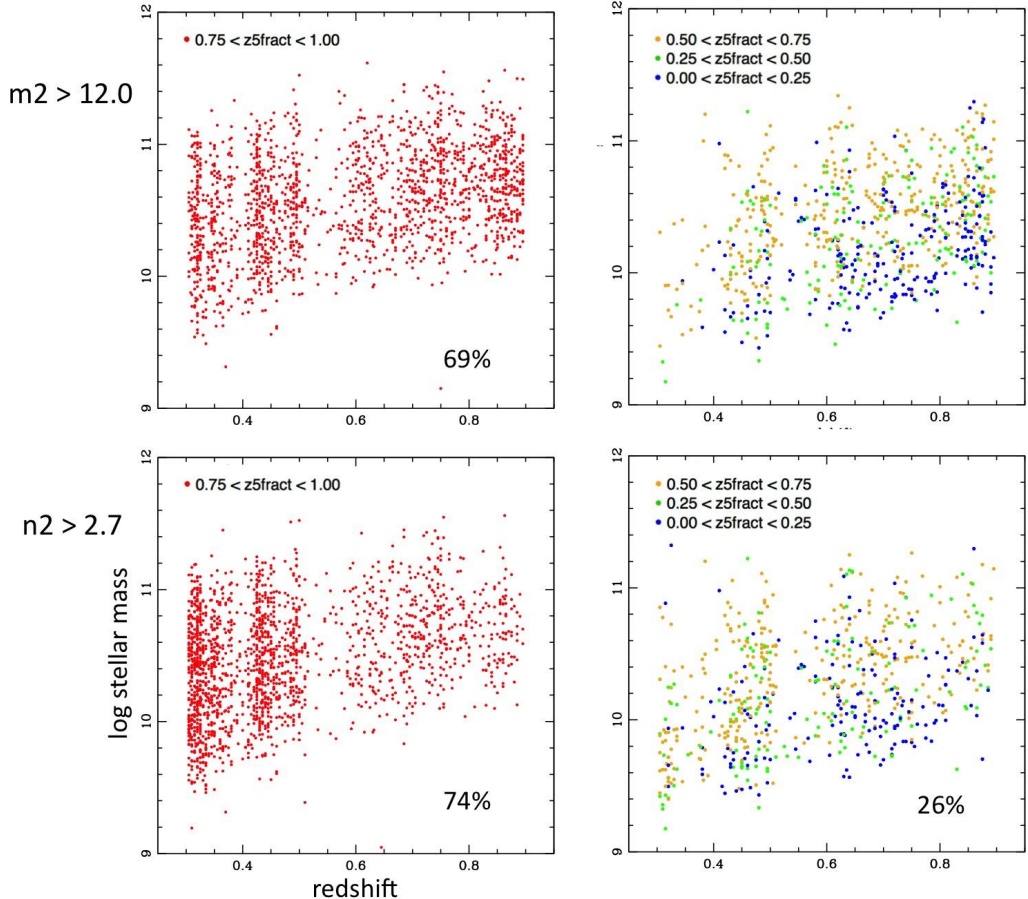}
}

\caption{Same as Figure 7, the fraction of mass produced in the first $\sim$5 Gyr, for the highest-density regions of the sample.
The top panels show the 11\% of galaxies in the densest regions in terms of projected mass density.  Points are galaxies
within a circle of (comoving) 2 Mpc radius and $\Delta$z/(1$+$z) $<$ 0.03.  The 11\% of the sample shown here is
in regions with m2 $>$ 12 $\times 10^{10}$\,\Msun\ per co-moving Mpc$^2$.  The bottom panels show the 11\% of 
the sample in the densest regions according to number density, n2 $>$ 2.7 signifies number density of more than 2.7 galaxies 
within a 2 Mpc (co-moving) circle and $\Delta$z/(1$+$z)$<$0.03).  Consistent with expectations, the sample in Figure 8 is richer 
in the  galaxies with the largest fraction of stars formed in the first $\sim$5 Gyr (red points on the left panels), \emph{z5fract} 
$>$ 0.75.  The 69\% and 74\% fractions of these galaxies are significantly larger than the 62\% for the complete sample (see Figure 7). 
The large-scale structure of the volume is shown more  clearly than for the full sample, as is expected with the larger percentage of 
these older, more massive galaxies.  Also as expected, the comparatively less-massive galaxies with SFHs that are more spread out over 
cosmic time, or even young (\emph{z5fract} $<0.5$), are less concentrated in the denser regions.  However, despite the well-known greater 
fractions of  ``passive" galaxies in these denser regions, the similarity between Figures 7 and 8 in the distribution of each of the four 
\emph{z5fract} categories indicates that the local environment is not driving SFHs, but  only regulating their fractional occurrence, as 
described in the text.
}

\end{figure*}


\begin{figure*}[t]

\centerline{
\includegraphics[width=6.5in, angle=0]{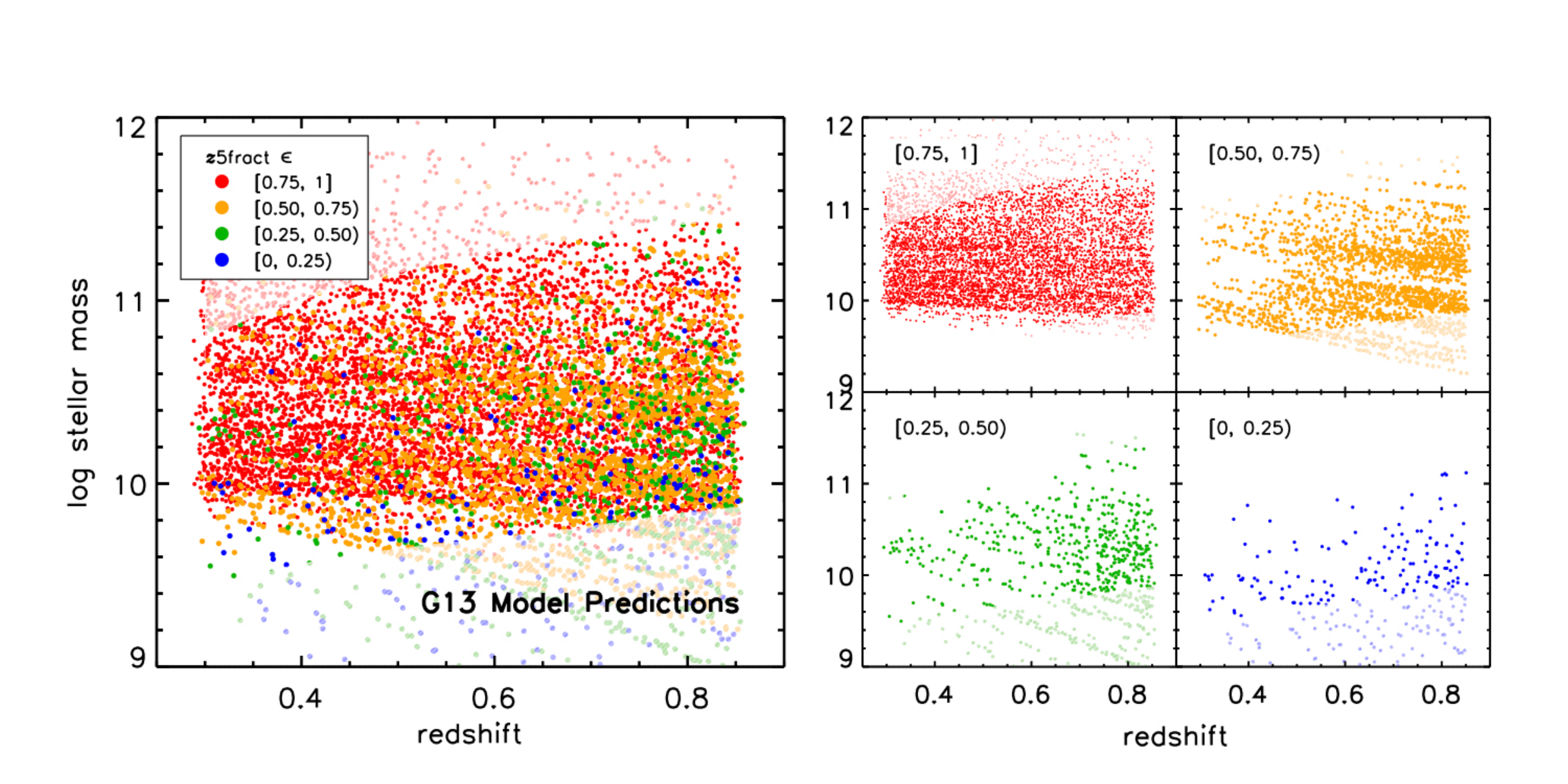}
}

\label{fig:G13D16}

\caption{The $M_{*}$/$z$/{\it z5fract} plane as predicted by the G13 model. The coloring is identical to Figure 7, but points here 
show ``reobservations'' of the suite of G13 lognormal SFHs at different times. The model and data trends are quite similar: (1) great 
SFH diversity exists at $M_{*}\lesssim10^{11}\, M_{\odot}$ and $z\gtrsim0.3$; (2) there is broad overlap between different 
{\it z5fract} classes; but (3) there is a clear preference for ``late bloomers'' (blue/green points) to lie at lower masses than ``early 
formers" (red/orange) at any/all epochs. ``Late bloomers'' also drop in abundance at later times. The shaded bottom wedge denotes 
one-third of the CSI stellar mass completeness threshold, corresponding roughly to the limit of the data. The top shaded envelope shows 
$dV/dz$---the change in survey volume as a function of redshift---scaled to match the observed edge.}

\end{figure*}


\begin{figure*}[t]

\centerline{
\includegraphics[width=6.3in, angle=0]{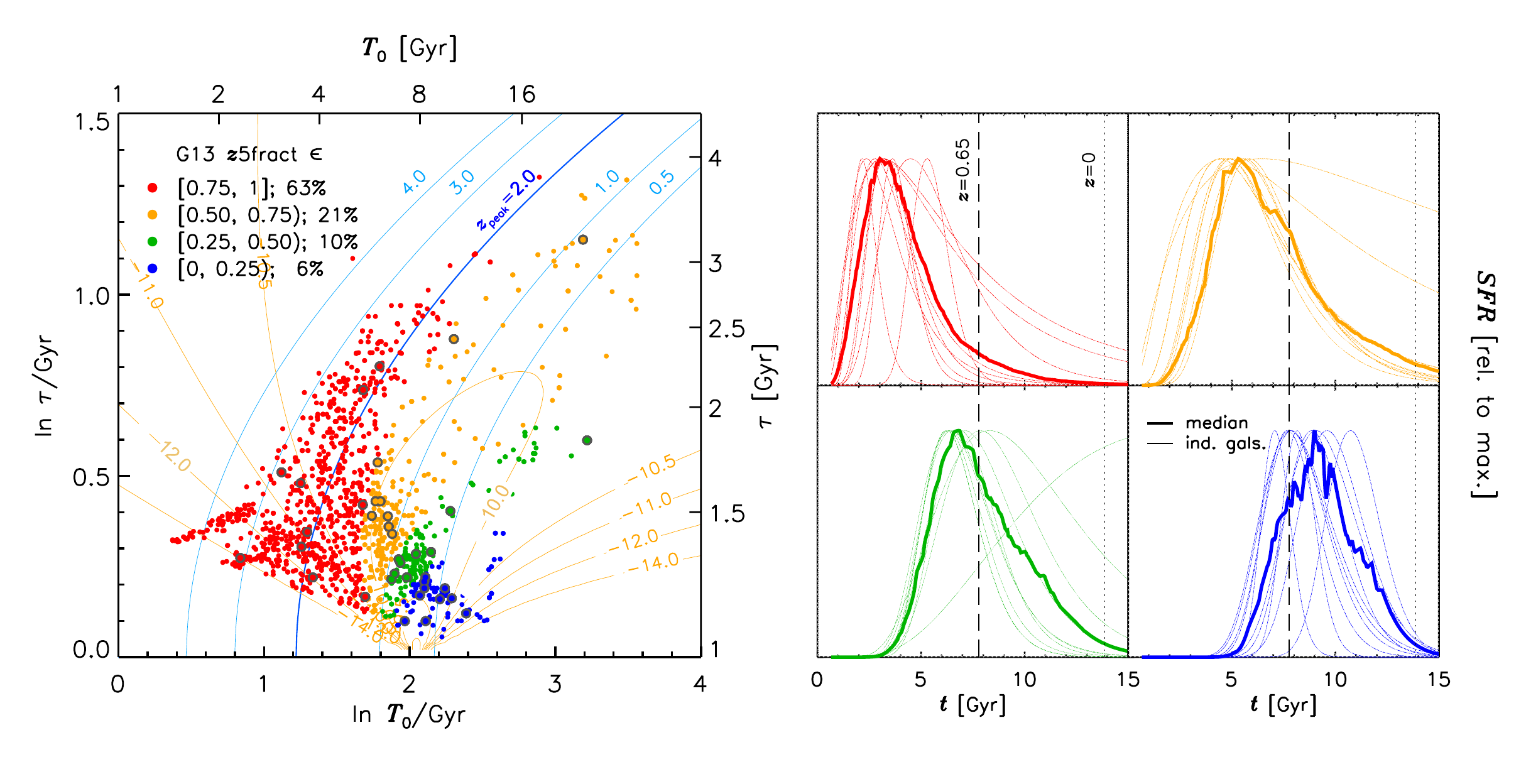}
}

\caption{Left: The G13 $(t_{0},\tau)$ plane color-coded by \emph{G13\,z5fract} at $z=0.65$. Here, \emph{G13/,z5fract} reflects stellar mass
formed prior to 1.7 Gyr before observation; that is, $t_{\rm obs}-\langle\tau\rangle$, a more natural way to identify ``old'' and ``young''
galaxies in the G13 scheme (Equation \ref{eq:G13z5fract}). Class fractions---with associated min/max ranges---are all within $\pm10\%$
of the observed values.  Blue lines show loci of constant SFH peak time (at listed $z$), and orange lines show loci of constant log sSFR (in yr$^{-1}$).
Note that the sSFR contours move over time while the points stay fixed (see Figure 9 of Gladders \etal\ 2013, which shows $z\approx0$ contours), 
reflecting the fact that cosmic SF is supported by different parts of $(t_0,\tau)$ space at different times.  At $z=0.65$, objects inside the loop 
at ${\rm log\,sSFR}=-10\,\,\rm{yr^{-1}}$ are the dominant contributors to SFR density, while at earlier times objects to the right  
dominated (falling SFHs at this epoch), and later, objects to the left {\it will} dominate (rising SFHs at this epoch). The left panel explicitly 
links this back to the {\it z5fract} classes.  Right: the median G13 SFH for each class (thick lines) and those for 10 representative 
members (thin lines; from points circled in black at \emph{\it left}). As anticipated from Figures 1 and 7, red objects have rapidly 
falling SFHs; amber objects have slowly falling SFHs; green objects are being observed near the peak of their SFHs; and blue objects 
are still rising. Hence, the $(t_{0},\tau)$ plane meaningfully characterizes the trends derived from the data, especially once the 
different timescales denoting ``youth'' are accounted for.}

\label{fig:t0Tau}

\end{figure*}

From one standpoint, showing the data of Figure 8 is one way to answer questions about the content of previous diagrams in this paper, 
specifically, whether \emph{satellites} and \emph{centrals} need to be treated separately, because---in the paradigm---many galaxies in 
dense regions are satellites, while most of the galaxies in sparse regions are centrals.  In fact, though, Figure 7 and Figure 8 are so similar 
that removing the latter's points from Figure 7 would make no discernible change.  In other words, the analysis we have made showing 
the diversity of SFHs does not show a strong environmental dependence. (It is very important to remember,  however, that there are no rich 
clusters of galaxies in the CSI XMM region.) 

However, a more fundamental point can be made about the way the different \emph{z5fract} subsets in Figures 7 and 8 are distributed in mass: 
\emph{they appear the same}.  Figure 8 has larger fractions of  ``old'' galaxies than the complete sample, 69\% and 74\% compared to the 
full sample's 62\%.  This is a global average, of course: looking at the dense regions at $z\sim0.33$ and $z\sim0.45$, the proportion of  ``old'' 
galaxies is larger still.  This is a natural expression of the stronger clustering of more massive, old galaxies: these trace the large-scale structure 
more clearly than previous diagrams.  Nevertheless, the distributions in mass for each subset appear indistinguishable, even in these densest 
regions, and even though their proportions are different.  In other words, there are different proportions of SFHs, but when they occur, their 
properties are the same.  This is a familiar finding for studies of galaxy properties in different environments.  In particular, the distributions of standard morphological types look the same and share the same properties in regions of high and low density, even though the proportions 
of these types are very different.

Our brief treatment of this question is insufficient, but this quick look indicates that it is unlikely that the proposed satellite/central dichotomy  
plays a major role in shaping SFHs of galaxies in the most common environments, from relatively isolated galaxies to groups of tens of galaxies.

\section{Consistency of CSI SFH\MakeLowercase{s} with the G13 Lognormal SFH Model}

As Figures 7 and 8 show, genuinely young galaxies are found over a wide mass range, $3\times10^{9}$--$10^{11}\,M_{\odot}$, and, 
though their abundances and masses decrease after $z\sim0.4$, they are present at all epochs this study has probed.

The large spreads of masses and times at which young systems are found is the ``smoking gun'' of the kind of SFH diversity that 
distinguishes the G13 and Kelson (2014b) SFH paradigms from more conformal pictures, wherein physical inferences rely on the 
assumption that galaxies of a given stellar mass can be well described by a common SFH (plus a quenching event). The data clearly 
suggest that a substantial amount of astrophysics cannot be captured by such approaches.

An examination of the model predictions of G13 for how Figure 7 {\it should} look helps us understand these issues.  In this 
exercise, the data used are those of the G13 study, but they are here sampled in a manner similar to how galaxies were treated in
this study, that is, in terms of the ratio of stellar mass produced early compared to the total stellar mass at the epoch of observation.  
We show in Figure 9 a reproduction of Figure 7 based on the suite of G13 lognormal SFHs: the points are colored identically in the 
two figures, which are also based on the same definition of \emph{z5fract} (see Section 4). The shaded regions echo the completeness 
and volumetric effects evident in the data, although we leave them semitransparent to show the full range of the model  measurements.

As shown in the large (left) panel of Figure 9, G13 exhibits all of the qualitative trends seen in the data: there is a broad diversity in SFHs at 
all stellar masses and epochs, such that ``young'' galaxies (\emph{z5fract} $< 0.5$) occupy most of the diagram but also display a tendency 
to lower masses. 

The subpanels in Figure 9 (right) make the above points slightly more quantitatively. Here, we see that G13 shows that the 
``youngest'' systems (\emph{z5fract}$<0.25)$, blue points) reach $10^{11}$\,\Msun, but mainly at the earliest epochs probed. 
Very few of these systems appear above $10^{10.5}$\,\Msun\ at $z=0.3$, just as is seen in the data. The green points reach higher 
masses at slightly later times, on up to the oldest galaxies (red points), which span all masses above $10^{10}$\,Msun\ at all 
epochs.\footnote{Note that $10^{10}$\,\Msun\ is the $z=0$ mass limit of the G13 model.}

Such agreement is reassuring: it suggests that the G13 model captures or encodes at least some of the relevant astrophysical processes 
that more conformal approaches average out. However, once we move beyond the above metrics, the G13 model results begin to diverge 
quantitatively from the data.  Specifically, in its current form, G13 does not produce the right \emph{fractions} of the \emph{z5fract} classes: 
while, at $z=0.65$---the median redshift of the sample---the data reveal abundances of 56\%, 18\%, 10\%, and 16\% for the red, amber, 
green, and blue objects, respectively, G13 yields instead 81\%, 13\%, 3\%, and 2\% ($\pm1$--2\%).

Yet, this discrepancy is easy to understand: while 1 Gyr is a timescale on which spectral features change---and thus a natural timescale 
to use to \emph{assess} SFHs---it is \emph{not} a \emph{natural} timescale.  For example, in the Kelson (2014b) scheme, SFHs fluctuate 
on all timescales, but each history has ``memory" over a Hubble time.  Similarly, in the G13 scheme, SFHs fluctuate on $\tau$-like timescales---ranging 
from one to many gigayears---and are diversified from each other also over a Hubble time.

As such, instead of comparing G13 to the CSI data by matching measurement definitions, we can compare the two by matching definitions 
of {\it youth}. To do this, we redefine  \emph{G13\,z5fract} as
\begin{equation}
	\emph{G13\,z5fract} \equiv \frac{M_{*}(t_{\rm obs} - \langle\tau\rangle)}{M_{*}(t_{\rm obs})},
\label{eq:G13z5fract}
\end{equation}
where $\langle\tau\rangle = 1.7$ Gyr is the mean $\tau$ value for the G13 SFHs. Here, \emph{z5fract} becomes the fraction of mass 
formed before most SFHs can have changed from a low to high or high to low state of star formation.

With this modification, the galaxy fractions in the \emph{G13\,z5fract} classes are close to what is observed: 63\%, 21\%, 10\%, and 
6\% ($\pm1$--3\%) again compared to 56\%, 18\%, 10\%, and 16\% in the data (high to low \emph{z5fract}).

So, what can we learn from this more physically aligned model classification?

Figure 10 (left) shows the G13 $(t_{0},\tau)$ plane---the natural description of galaxies in this model---with all SFHs color-coded 
by their \emph{G13\,z5fract} values  (based on $\langle\tau\rangle$) at $z=0.65$.\footnote{Recall that galaxies do not move in this plane 
in the G13 framework, so this is a ``stationary frame'' in which to analyze the model SFHs.} As anticipated (see Sections 2 and 4), SFH 
stratification is mainly based on peak time, demarcated by the blue curved lines: high \emph{G13\,z5fract} objects have early-peaking SFHs 
($z_{\rm peak}\gtrsim1.5$), with peak times moving to later epochs with lower \emph{G13\,z5fract} objects.  Indeed, as revealed by a look 
at the example SFHs for each class shown at \emph{right}, the breakdown is nearly exactly as stated in Section 4 based on the data: while 
red SFHs are falling rapidly, amber objects are falling more gently, green objects are being observed near their peak of star formation, 
and blue ones are still on the rise, peaking sometime between $z=0.7$ and today.\footnote{An examination of the curves of constant sSFR is 
revealing in this context. At $z=0.65$, objects inside the loop at ${\rm log\,sSFR}=-10\,\rm{yr^{-1}}$ dominate the cosmic SFR density. 
They are not at all a homogenous set, but rather include galaxies from three {\it z5fract} classes.  Moreover, they have totally different 
SFHs than objects that dominated cosmic SF at higher $z$ (points to the left of the loop; falling SFHs), and those that {\it will} do so 
at lower $z$ (points to the right of it; rising SFHs). This fact highlights how ``young'' galaxies observed at any epoch cannot be taken to 
reflect analogs of the progenitors of older galaxies observed at the same epoch. Moreover, ``old''---in the sense of having built a lot of 
mass long ago---does not  always mean ``dead or dying.''}

Yet, the subtleties expressed within each class are revealing beyond this. For example, though the red objects are ``old''---and indeed 
will nearly all be dead by today (dotted vertical lines in right-hand panels)---many are still actively forming stars at the epoch of 
observation (dashed vertical lines). This belies a key point: \emph{galaxies with constant SFHs will fall into the highest G13 
z5fract bin at these epochs.}  Hence, {\it old does not mean dead}. (This point is clarified by noticing that many of the 
red points lie between the orange curves of constant $\log {\rm sSFR/yr^{-1}}>-11$ in Figure 10, left.) 

Furthermore, looking at the amber curves, we see that even some systems with high \emph{G13\,z5fracts} ($>$50\%) can have 
\emph{rising} SFHs (or relatively late starts).  This suggests that the current practice of modeling SFHs---not only for high-redshift 
galaxies, but even intermediate-redshift galaxies---with exponentially \emph{declining} SFHs can often be a mistake and serves as a 
``yellow flag'' to our intuitions.   Such are the effects induced by a two-parameter (or zero-parameter, in the case of Kelson 2014b) 
SFH model.

To the extent that those frameworks describe reality---which we think is substantial (Gladders 2013; Kelson 2014b; Abramson 2015; 
Abramson \etal\ 2016)---the above results thus suggest that identifying \emph{what} determines a galaxy's location in $(t_{0},\tau)$ 
space---or a similar plane to be defined for the Kelson model---is paramount. That is, even knowing a galaxy is ``old''  is not enough to pin 
down its SFH; that is, there is no ``typical'' history, as our discussion of the red and amber points highlights.  Additional physics could 
have \emph{substantially} influenced---not merely perturbed---some members of a given class more than others, and these are 
perhaps \emph{the} critical processes to characterize. 

The data in this study cannot uniquely reveal such causal mechanisms, but we can combine our discussion of environment in Section 4
with the G13 model results to come up with a hypothesis, as follows.

The fractions of the \emph{z5fract} classes change in denser regions (Figure 8), but not the overall trends (Figure 7).  This suggests a scenario 
in which late-time environmental effects (e.g., stripping, strangulation) simply do not play a major role. If they did, the \emph{z5fract} patterns 
in Figure 8 would be sensitive to the (diverse) infall histories of groups and the subsequent (chaotic) existences of galaxies ``postacquisition.'' 
That they---and a host of structural (e.g., Allen \etal\ 2016; Morishita \etal\ 2016) and star formation-related-properties (e.g., Peng \etal\ 2010; 
Newman \etal\ 2014; Vulcani \etal\ 2015b)---appear \emph{not} to be sensitive is evidence that galaxy histories are \emph{correlated with} the 
history of their environment, but not \emph{causally connected} to their presence in it at any given time. That is, most of the interesting physics 
controlling SFHs must be ``baked in'' at the \emph{start} of halo growth, that is, very early on.

This is the scenario presented by Dressler (1980). As discussed by Abramson \etal\ (2016), because each galaxy is assigned a 
$(t_{0},\tau)$ coordinate at birth in the G13 model, it is {\it also} the scenario that framework naturally describes. In either, galaxies 
in (rare) isolated, high-density halos {\it can} have the same early-peaking SFHs as those in denser regions, but because a common 
initial ``low-frequency'' mode will push the galaxy-scale peaks riding on top of it to collapse earlier, such objects will be significantly 
biased toward what will become groups and clusters. Environment thus determines a global preference for $(t_0,\tau)$---that is, low 
values---but does not fundamentally restrict them. In this way, though it need not have been, a scenario in which initial conditions 
largely determine SFHs is fully consistent with the new measurements presented here. 


\section{Summary and Conclusions}

Our principle conclusions are as follows:

\begin{enumerate}

\item{Coarse SFHs from the CSI Survey exhibit a diversity of SFHs that are nonconformal.  That is, they are neither 
replicas nor appear organized by the behavior of macroscopic/population-level scaling laws.  Rather, a two-parameter description, such 
as an SFR that is lognormal in time---with its double timescales of \T0 and $\tau$ (Figures 9 and 10)---seems to be the minimum required 
to reproduce the diversity of SFHs found in our study of the CSI Survey data for galaxies of stellar mass log\,M/\Msun  \,= $9-12$ 
and $0.3 < z < 0.9$.} 

\item{Our analysis of the CSI Survey data demonstrates that broad-wavelength coverage and accurate spectrophotometry can 
constrain the populations of stars formed over ``natural'' timescales of stellar evolution, of $10^{10}, 10^9$, and $10^8$ years.  
SFRs calculated from a sophisticated fitting of spectral energy distributions, based on well-understood signatures of stellar 
populations, are able to distinguish galaxies that formed all their stars early---in a few billion years---from those that formed 
stars continuously, and from those that formed most of their stars late---after $z=1$.}

\item{We confirm predictions by O13 and modeling by G13 by identifying a substantial population of genuinely young 
galaxies, those that formed most of their stellar mass after $z=1$ and within 1 Gyr of the epoch of observation.}  

\item{We demonstrate through duplicate measurements of galaxy stellar mass that a parameter \emph{z5fract}---the fraction 
of total stellar mass that was formed in the epoch starting at $z=5$ and lasting $\sim$5 Gyr---is a simple but robust indicator of 
SFHs.  We quantify the relative proportions of diverse SFHs through \emph{z5fract} and show that, at $z\sim0.6$ and 
log M$_* \sim 10$, genuinely young galaxies---those with greater stellar mass made in the final gigayear than in the time 
prior---are $\sim$50\% of the population, and that---though they are rare---there are young galaxies with masses of the 
Milky Way, $4 \times 10^{10}$\,\Msun, and higher.}

\item{The different forms of SFHs are functions of total mass, but there is a large diversity at any given mass and, conversely, a wide
range of mass ($\sim$1.0 dex) over which a particular SFH can be found.  Assuming that stellar mass is strongly coupled to dark-halo 
mass, this means that galaxies with the same halo mass at any given epoch can host very different SFHs.  This suggests another 
parameter, in addition to halo \emph{mass}, that controls when star formation begins and how it progresses.  This could be a property 
of the halo, for example, its density or turnaround time, or of baryonic physics, such as star-formation or black hole growth feedback.  
These factors could drive the ``efficiency" of turning baryonic mass into stars that is offered as an explanation of why the halo mass 
function and the stellar mass function of galaxies do not ``track.''}

\item{Our sample is dominated by ``field" galaxies, which means isolated galaxies or those in moderate-mass groups. 
The most massive galaxies, and those galaxies most exposed to `environment effects,' are not represented in the 
present study because the CSI Survey includes no rich clusters.  That being said, the SFHs we present account for $\sim$95\% 
of all galaxies with masses larger than  10$^9$\,\Msun.  The conformal approach to SFHs explicitly includes ``quenching 
mechanisms" to alter the course of star formation, specifically, to stop it, but only a small fraction of our sample are good 
candidates for ``mass quenching" or ``environmental quenching,'' two popular generic concepts.  Nevertheless, a very large 
fraction of our 22,494 galaxy sample have evolved from SFRs of tens of \MsunYr\ early in their lifetime to $\ls$1 \MsunYr\ by 
the present epoch.  Our data therefore suggest that natural processes working on the timescale of  a Hubble time are able to 
regulate rising and then falling star formation without discrete, short-timescale quenching events.  The picture that emerges is one 
of predestined paths for galaxies, perhaps modulated by mergers in their early history, with a ``clock'' and as-yet unidentified 
intrinsic processes that propel a galaxy---\emph{sooner or later}---to a state where halo gas is insufficient or in a unusuable 
state for continued star formation.}

\item{These and other data support a picture where SFHs are generally locked in at an early epoch, determined by specific
properties of the environment or the galaxy itself.  Such built-in trajectories can provide an alternative explanation
for supposedly environment-related correlations, like galaxy morphology with local density or the satellite/central
paradigm, as expressions of \emph{initial conditions}.  Dark-matter halos must play a role in this process, but a simple
correlation of SFH and halo mass is clearly incompatible with the diverse histories we present here.  We look to 
theoretical work to explore halo or baryonic properties that are responsible for this manifold variety of SFHs.}

\end{enumerate}

\section{Acknowledgments}

The authors thank the scientists and staff of the Las Campanas Observatories for their dedicated and effective 
support over the many nights of the CSI Survey.  B.V. acknowledges the support from an Australian Research Council 
Discovery Early Career Researcher Award (PD0028506).

\end{document}